# Experimental Observations of Nuclear Activity in Deuterated Materials Subjected to a Low-Energy Photon Beam

Bruce M. Steinetz and Theresa L. Benyo
National Aeronautics and Space Administration
Glenn Research Center
Cleveland, Ohio 44135

Vladimir Pines and Marianna Pines
PineSci Consulting
Avon Lake, Ohio 44012

Lawrence P. Forsley
JWK Corporation
Annandale, Virginia 22003

Paul A. Westmeyer
National Aeronautics and Space Administration
Washington, DC 20546

Arnon Chait
National Aeronautics and Space Administration
Glenn Research Center
Cleveland, Ohio 44135

Michael D. Becks
Vantage Partners, LLC
Brook Park, Ohio 44142

Richard E. Martin
Cleveland State University
Cleveland, Ohio 44115

Robert C. Hendricks
National Aeronautics and Space Administration
Glenn Research Center
Cleveland, Ohio 44135

Nicholas Penney
Ohio Aerospace Institute
Brook Park, Ohio 44142

Annette M. Marsolais and Tracy R. Kamm
Vantage Partners, LLC
Brook Park, Ohio 44142

## Summary

Exposure of highly deuterated materials to a low-energy (nom. 2 MeV) photon beam resulted in nuclear activity of both the parent metals of hafnium and erbium and a witness material (molybdenum) mixed with the reactants. Gamma spectral analysis of all deuterated materials, $ErD_{2.8}+C_{36}D_{74}+Mo$ and $HfD_2+C_{36}D_{74}+Mo$, showed that nuclear processes had occurred as shown by unique gamma signatures. For the deuterated erbium specimens, posttest gamma spectra showed evidence of radioisotopes of erbium ($^{163}Er$ and $^{171}Er$) and of molybdenum ($^{99}Mo$ and $^{101}Mo$) and by beta decay, technetium ($^{99m}Tc$ and $^{101}Tc$). For the deuterated hafnium specimens, posttest gamma spectra showed evidence of radioisotopes of hafnium ($^{180m}Hf$ and $^{181}Hf$) and molybdenum ($^{99}Mo$ and $^{101}Mo$), and by beta decay, technetium ($^{99m}Tc$ and $^{101}Tc$). In contrast, when either the hydrogenated or non-gas-loaded erbium or hafnium materials were exposed to the gamma flux, the gamma spectra revealed no new isotopes. Neutron activation materials showed evidence of thermal and epithermal neutrons. CR–39 solid-state nuclear track detectors showed evidence of fast neutrons with energies between 1.4 and 2.5 MeV and several instances of triple tracks, indicating >10 MeV neutrons. Further study is required to determine the mechanism causing the nuclear activity.

## 1.0 Introduction

Power systems for deep-space and planetary missions that cannot rely on solar power have for the most part exploited heat sources based on $^{238}Pu$. For instance, the Mars Science Laboratory's Radioisotope Thermoelectric Generator provides nominally 2 kW of thermal power, which is then converted to electric power (Ref. 1). Limited available quantities and the expenses associated with the production and safe handling of this isotope of plutonium have motivated the NASA Glenn Research Center to seek alternative sources of energy with long operating lives of a decade or more.

---

[1]This document reports preliminary findings intended to solicit comments and ideas from the technical community and is subject to revision as analysis proceeds.

[2]Trade names and trademarks are used in this report for identification only. Their usage does not constitute an official endorsement, either expressed or implied, by the National Aeronautics and Space Administration.

The ideal energy source would be light and compact, be maintenance free, would deliver gigajoule levels of energy over a decade in operation, would not require enriched materials, and could be actively controlled. To date, nuclear-based power generation is the only known technology with the required power or energy density (power or energy per unit mass) that could continuously operate for an extended period. The Advanced Energy Conversion (AEC) effort is exploring new methods for initiating nuclear activity that could potentially provide alternative routes for achieving the above attributes.

Deuterium, with one proton and one neutron, has been used as a nuclear material for many decades for applications ranging from inertial confinement fusion (ICF) reactors through neutron generators. Deuterium owes some of its key properties to its single positive nuclear charge and therefore allows for the lowest barrier for tunneling the electrostatic barrier for nuclear fusion. Deuterium fusion, however, generally requires at least 10 to 15 keV in kinetic energy (corresponding to over 100 million Kelvin) to raise the probability of tunneling to occur. Despite intense work over many decades, to date no "hot fusion" nuclear reactor has been demonstrated with a coefficient of performance greater than 1 (net positive power output) for sustained periods. Deuterium is available in nature, and it is separated from seawater where it has natural abundance of 0.0156%. Deuterium is stable (Ref. 2). From a nuclear standpoint, it has a low binding energy, though still quite significant at 2225 keV (with eV precision) (Ref. 3). Indeed, deuterium has been the focus of attention of many attempts over the years to exploit one or more of its unique properties to achieve alternative forms of nuclear activity.

If novel nuclear reactions involving deuterium were to occur, they would follow conventional rate relations that describe all nuclear processes. Essentially, the rate of nuclear processes is proportional to the product of the respective number densities of the reactants and the cross section, as well as to other parameters (e.g., the so-called Gamow factor describing the probability of two particles to overcome the electrostatic repulsion barrier in nuclear fusion). Therefore, for deuterium to participate in a reaction, it would be advantageous to bring its number density to near solid-state condition. In nature, such conditions are possible using deuterated metals, where the atomic ratio of deuterium to the host metal can be made greater than unity with the proper loading conditions (Ref. 4). Moreover, many hydrides can be loaded and do maintain such high number density after loading. They can be transferred to the reaction chamber under ambient temperatures and pressures without appreciable loss of loading. There are also many organic materials that have both hydrogenated and deuterated forms. For example, in the past, ICF targets have used deuterated polyethylene (Ref. 5) as a source of deuterium at near-solid-deuterium densities. Other methods to achieve high D-number density and cause novel reactions include using high $D_2$ gas pressures. Achieving high-number-density deuterium in this fashion can result in high mobility and other reactions of interest. For instance, Didyk et al. showed large heat release and novel material transmutations using high-pressure deuterium (3 kBar) in a linear accelerator (LINAC) beam (Ref. 6).

The findings of the current work as will be described may have applications beyond power production, including creating valuable radioisotopes. Processes are being sought to find effective ways of producing medical isotopes such as $^{99}$Mo and its decay product, $^{99m}$Tc, the most widely used medical radioisotope. The $^{99m}$Tc is used as a radioactive tracer element in more than 40 million medical diagnostic procedures each year, including heart stress tests and bone scans. The Nuclear Energy Agency (NEA) Steering Committee for Nuclear Energy established the High-level Group on the Security of Supply of Medical Radioisotopes (HLG–MR) in April 2009. The main objective of the HLG–MR was to strengthen the reliability of $^{99}$Mo and $^{99m}$Tc supply in the short, medium, and long term (Ref. 7). The issue of finding a replacement for production of the $^{99}$Mo isotope has become acute for the United States because its main supplier of $^{99}$Mo is the National Research Universal reactor (NRU, a highly enriched uranium reactor). This reactor at Chalk River, Ontario, operated by Canadian Nuclear Laboratorie, is scheduled for closing. The reactor was scheduled to shut down in 2016 but because of the lack of any commercially viable substitute process, plant decommission has been delayed until 2018 (Ref. 8).

This study represents partial results from a larger study, which began using a higher 6-MeV LINAC with deuterated materials. Above the photodissociation energy of deuterium of 2.226 MeV, it is expected to observe nuclear signatures consistent with neutron activation, kinetic heating, and other processes. In the quest for a smaller, lighter weight overall system design, and to investigate how low of energy would still result in nuclear activation, the team explored lower MeV exposures. The present study explores and reports reactions of metals and hydrocarbons containing deuterium at high number density plus witness materials, subjected to moderate-energy photons at energies determined to be less than the deuteron photodissociation threshold, using the water tank ionization chamber method.

# 2.0 Experimental Description

The experimental protocol is based, in general, on a case-control research methodology and on pre- and postexposure comparative evaluation of the materials. Multiple and orthogonal techniques were used to explore nuclear activation, formation, and decay of active radioisotopes, and to bound neutron energies. The materials and instrumentation utilized are covered in separate sections herein. Generally, the experimental protocol followed the procedure outlined as follows:

- Creation of documentation packages, called travelers, for tracking specimen history for analysis and testing
- Pretest materials characterization of as-received materials: (1) pretest nuclear diagnostics including alpha/beta and beta scintillation, (2) gamma spectroscopy establishing baseline signatures, and (3) infrared (IR) spectroscopy defining D-to-H ratios in the deuterated hydrocarbons for quality control
- Loading of reactants into test fixture per test plan specification
- Alignment of test fixture in LINAC beam
- Establishment of test laboratory and data system with LINAC in ready-to-run condition
- Exposure of test specimens to consistent beam conditions and time intervals
- Protocol for safe entry into bunker after test
- Unloading of reactants from test fixture
- Posttest nuclear diagnostics of materials: alpha/beta and beta scintillation and gamma spectroscopy
- Data management and archiving of digital and hard copy records
- Posttest tracking and storage of samples

## 2.1 Test Matrix and Samples

The samples exposed in this study were created from prepared batches of either deuterium- or hydrogen-loaded or bare (no-load) erbium or hafnium metal material. Molybdenum (no-load) was also included as a witness material as a measure of activation during the tests. The samples were placed into glass vials and subsequently positioned in the exposure path of a modified industrial LINAC. The vials were placed in a holder that could hold either two or four reactant vials at a time at close distance to the LINAC tungsten-braking target (Fig. 1).

### 2.1.1 Deuterated Materials

Natural-abundance base materials erbium and hafnium were deuterated or hydrogenated by gas loading using appropriate pressure, temperature, and time protocols. Metal purity was as follows: Er (99%), Hf (99.9%), and Mo (99.9%). Based on their respective phase diagrams, deuterated metals were chosen that would retain their "D-loading" under ambient conditions to simplify transfer into sample vials for beam exposure. The sample mass change (accuracy $\pm$5%) from before to after gas loading was used to determine the D or H loading of the sample materials. Note 99.999% ultra-high-purity gas was used to deuterate or hydrogenate the samples. The mass change was then used to calculate the ratio of the gas load atoms per metal host atom. The materials were blended and mixed with a precleaned glass mortar and pestle and when fully mixed, poured into a new 7-ml glass vial. Figure 2 is an end view of four vials in the specimen holder prior to mounting them in the LINAC beam. Table I presents the material masses in each sample mixture and also presents the exposure energy and times for each specimen. In addition to the deuterated base metals of $ErD_x$ and $HfD_2$, additional high-number-density deuterated paraffin ($C_{36}D_{74}$; with C–D/C–H ratio $\geq$97%) was mixed in with the molybdenum witness material. The term "witness material" is used here because the presence of this material provided insight into the types of nuclear reactions that occurred during the beam exposure as measured posttest in the gamma and beta counters.

### 2.1.2 Control Tests: Hydrogenated Materials

In addition to tests with deuterated materials, two types of control tests were performed. The first control used hydrogenated versions of the natural-abundance erbium and hafnium mixed with the corresponding H-paraffin (H-para, $C_{36}H_{74}$) and target molybdenum powder. Samples with the same nominal masses as the deuterated versions were exposed in glass vials at the same distance from the braking target. Two copies of the test articles were simultaneously exposed in the beam for both the erbium and hafnium material systems.

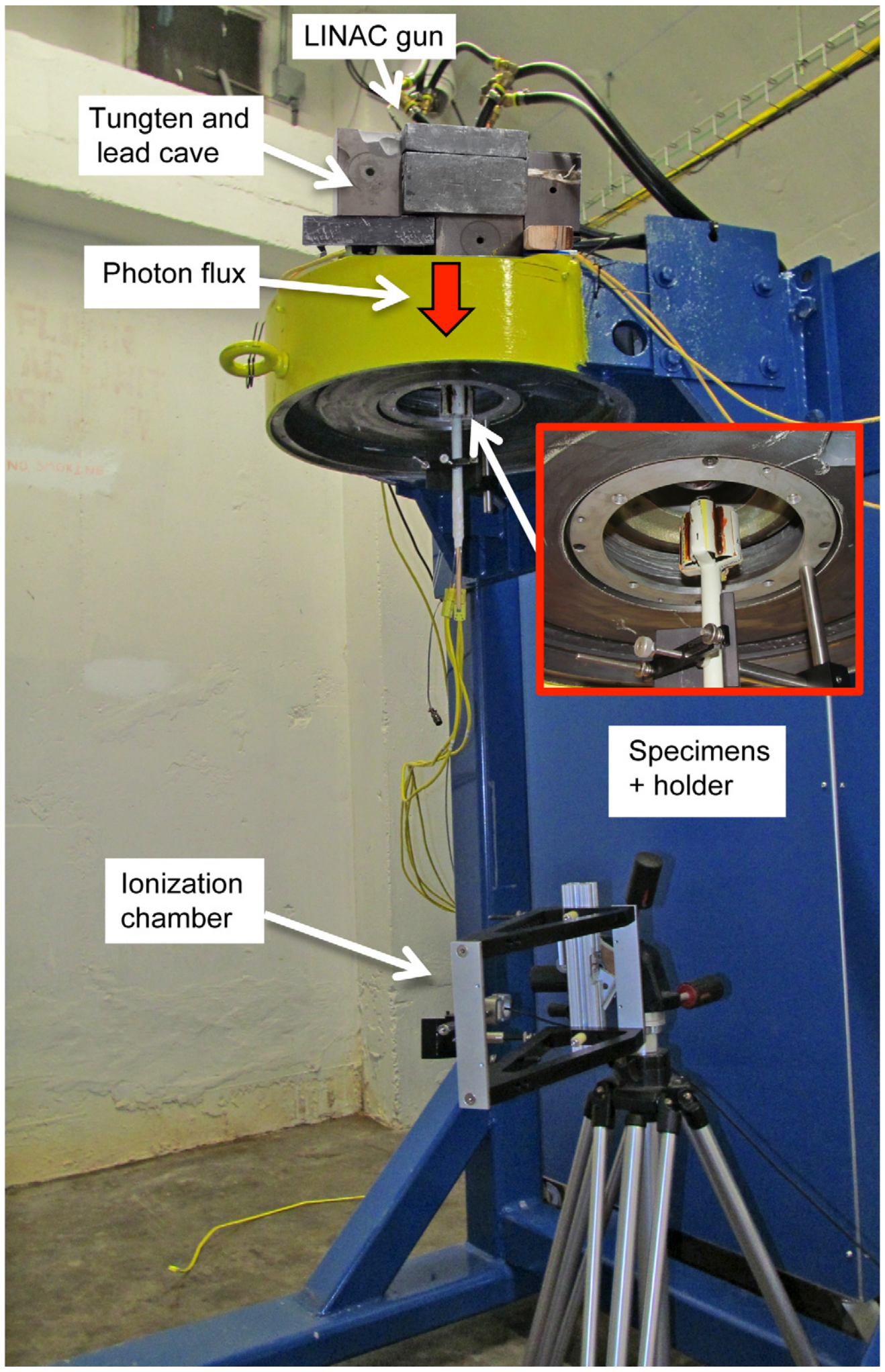

Figure 1.—Varian nom. 2-MeV LINAC system with specimens positioned just below braking target and ionization chamber radiation detector positioned 100 cm below target. A tungsten cave was added in later tests to shield neutron activation materials from the gamma exposure.

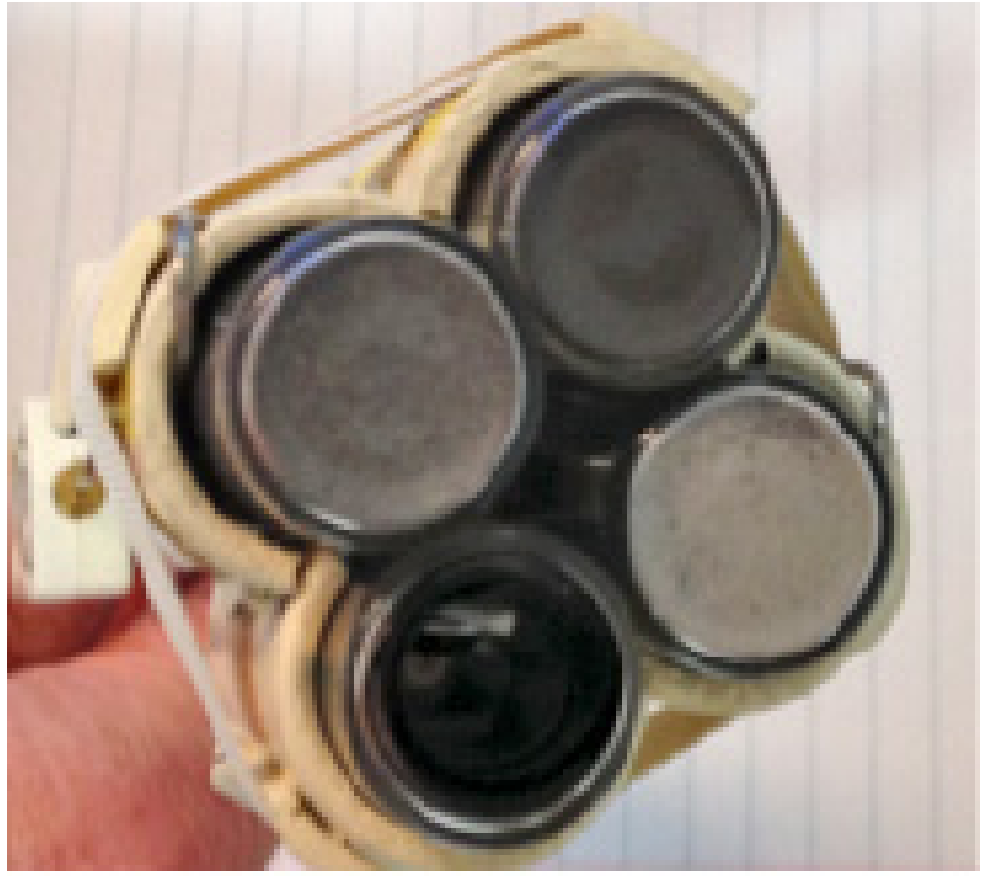

Figure 2.—End-on view of four test specimens in vials prior to LINAC beam exposure. Center of holding fixture aligned with beam centerline. Vial-to-vial (diagonal) distance nominally 2.5 cm (1 in.).

TABLE I.—MATERIALS EXPOSED TO THE HIGH-FLUX GAMMA ENERGY

| Test shot number | Duration, hr | Date run | Fuel 1 | | Fuel 2 | | Witness material | |
|---|---|---|---|---|---|---|---|---|
| | | | Material | Mass, g | Material | Mass, g | Material | Mass, g |
| $ErD_x$+$C_{36}D_{74}$+Mo | | | | | | | | |
| PGL 2028a | 4 | 28-Jul-15 | $C_{36}D_{74}$ | 2.0003 | $ErD_3$ | 6.0045 | Mo | 6.0013 |
| PGL 2029aa | 4 | 28-Jul-15 | $C_{36}D_{74}$ | 2.0003 | $ErD_3$ | 6.0018 | Mo | 6.0024 |
| PGL 2050 | 4 | 3-Aug-15 | $C_{36}D_{74}$ | 2.0000 | $ErD_{2.8}$ | 6.0008 | Mo | 6.0008 |
| PGL 2051 | 4 | 3-Aug-15 | $C_{36}D_{74}$ | 2.0001 | $ErD_{2.8}$ | 6.0000 | Mo | 6.0000 |
| PGL 2052 | 4 | 3-Aug-15 | $C_{36}D_{74}$ | 2.0001 | $ErD_{2.8}$ | 6.0005 | Mo | 6.0003 |
| PGL 2053 | 4 | 3-Aug-15 | $C_{36}D_{74}$ | 2.0000 | $ErD_{2.8}$ | 5.9928 | Mo | 6.0000 |
| $HfD_2$+$C_{36}D_{74}$+Mo | | | | | | | | |
| PGL 2034 | 4 | 29-Jul-15 | $C_{36}D_{74}$ | 1.99701 | $HfD_2$ | 5.99029 | Mo | 5.99029 |
| PGL 2035 | 4 | 29-Jul-15 | $C_{36}D_{74}$ | 1.99701 | $HfD_2$ | 5.99029 | Mo | 5.99029 |
| PGL 2068 | 4 | 6-Aug-15 | $C_{36}D_{74}$ | 2.0001 | $HfD_2$ | 6.213606 | Mo | 5.98744 |
| PGL 2069 | 4 | 6-Aug-15 | $C_{36}D_{74}$ | 2.0000 | $HfD_2$ | 6.213475 | Mo | 5.98731 |
| PGL 2070 | 4 | 6-Aug-15 | $C_{36}D_{74}$ | 2.0002 | $HfD_2$ | 6.214131 | Mo | 5.98795 |
| PGL 2071 | 4 | 6-Aug-15 | $C_{36}D_{74}$ | 2.0063 | $HfD_2$ | 6.232902 | Mo | 6.00603 |
| $ErH_3$+$C_{36}H_{74}$+Mo | | | | | | | | |
| PGL 2064 | 4 | 5-Aug-15 | $C_{36}H_{74}$ | 2.0009 | $ErH_3$ | 6.0012 | Mo | 6.0009 |
| PGL 2065 | 4 | 5-Aug-15 | $C_{36}H_{74}$ | 2.0009 | $ErH_3$ | 6.0012 | Mo | 6.0009 |
| $HfH_2$+$C_{36}H_{74}$+Mo | | | | | | | | |
| PGL 2077 | 4 | 7-Aug-15 | $C_{36}H_{74}$ | 2.0005 | $HfH_2$ | 6.0006 | Mo | 6.0005 |
| PGL 2078 | 4 | 7-Aug-15 | $C_{36}H_{74}$ | 2.0005 | $HfH_2$ | 6.0006 | Mo | 6.0005 |
| Er+Mo | | | | | | | | |
| PGL 2061 | 4 | 4-Aug-15 | None | N/A | Er, no load | 2.0818 | Mo | 2.0822 |
| Hf+Mo | | | | | | | | |
| PGL 2060 | 4 | 4-Aug-15 | None | N/A | Hf metal chunk, no load | 6.0017 | Mo | 6.0006 |
| $ErD_x$+ $C_{36}D_{74}$ +Mo | | | | | | | | |
| PGL 2150 | 6 | 30-Oct-15 | $C_{36}D_{74}$ | 1.9998 | $ErD_{2.8}$ | 6.000110 | Mo | 6.00011 |
| PGL 2151 | 6 | 30-Oct-15 | $C_{36}D_{74}$ | 2.0001 | $ErD_{2.8}$ | 6.000044 | Mo | 6.00004 |
| PGL 2152 | 6 | 30-Oct-15 | $C_{36}D_{74}$ | 2.0001 | $ErD_{2.8}$ | 6.000453 | Mo | 6.00026 |
| PGL 2153 | 6 | 30-Oct-15 | $C_{36}D_{74}$ | 2.0010 | $ErD_{2.8}$ | 5.996031 | Mo | 6.00298 |
| PGL 2194 | 6 | 23-Nov-15 | $C_{36}D_{74}$ | 2.0257 | $ErD_{2.71}$ | 5.995921 | Mo | 5.99530 |
| PGL 2195 | 6 | 23-Nov-15 | $C_{36}D_{74}$ | 2.0258 | $ErD_{2.71}$ | 5.996178 | Mo | 5.99555 |
| PGL 2196 | 6 | 23-Nov-15 | $C_{36}D_{74}$ | 2.0258 | $ErD_{2.71}$ | 5.996220 | Mo | 5.99560 |
| PGL 2197 | 6 | 23-Nov-15 | $C_{36}D_{74}$ | 2.0257 | $ErD_{2.71}$ | 5.995921 | Mo | 5.99530 |
| $HfD_2$+$C_{36}D_{74}$+Mo | | | | | | | | |
| PGL 2142 | 6 | 28-Oct-15 | $C_{36}D_{74}$ | 2.0159 | $HfD_2$ | 5.99410 | Mo | 5.99368 |
| PGL 2143 | 6 | 28-Oct-15 | $C_{36}D_{74}$ | 2.0156 | $HfD_2$ | 5.99325 | Mo | 5.99283 |
| PGL 2144 | 6 | 28-Oct-15 | $C_{36}D_{74}$ | 2.0157 | $HfD_2$ | 5.99350 | Mo | 5.99308 |
| PGL 2145 | 6 | 28-Oct-15 | $C_{36}D_{74}$ | 2.0155 | $HfD_2$ | 5.99273 | Mo | 5.99231 |
| PGL 2189 | 6 | 20-Nov-15 | $C_{36}D_{74}$ | 2.0134 | $HfD_2$ | 5.99691 | Mo | 5.99531 |
| PGL 2190 | 6 | 20-Nov-15 | $C_{36}D_{74}$ | 2.0135 | $HfD_2$ | 5.99734 | Mo | 5.99574 |
| PGL 2191 | 6 | 20-Nov-15 | $C_{36}D_{74}$ | 2.0136 | $HfD_2$ | 5.99746 | Mo | 5.99587 |
| PGL 2192 | 6 | 20-Nov-15 | $C_{36}D_{74}$ | 2.0119 | $HfD_2$ | 5.99241 | Mo | 5.99081 |

### 2.1.3 Control Tests: Bare Metals

The second control used standard natural-abundance erbium and hafnium materials mixed with molybdenum powder witness material with no gas loading. Samples with the same nominal masses as the deuterated versions were exposed in glass vials at the same distance from the braking target. A single copy of these test articles was exposed in the beam.

## 2.2 LINAC Photon Source

A LINAC Model LS200 manufactured by Varian was modified to expose the samples to photon energies of ≤2 MeV. A special dequeing circuit was added to the system, which reduced the beam energy from a maximum of 2.4 MeV to less than 2 MeV, as will be discussed in Section 2.2.2, End-Point Energy Measurement. This industrial LINAC allowed specimens to be positioned very close to the braking target. For this study, the specimens (Fig. 2) were positioned within approximately 7.4 mm (0.29 in.) from the exit plane of the tungsten-braking target. No flattening filters for beam energy were used in these tests. At 7.4 mm, it is estimated that the samples saw a radiation dosage of 2.4 E6 rad/min at the nominal 2-MeV beam energy. An ionization gauge (RadCal PN 10X6-0.6) radiation detector was set up below the test samples at the isocenter (~100 cm from the braking target) to monitor radiation levels emanating from the LINAC beam. The chamber measures charge buildup over a sampling period proportional to radiation levels impinging on the detector. The radiation level as well as reflected power, gun current (voltage), and target current (voltage) were recorded and used to monitor beam operation to ensure that the beam flux was not changing with time.

### 2.2.1 Calculated Photon Flux

Using the methods outlined in Reference 6, the bremsstrahlung photon flux was calculated over the energy range of 81.5 keV to 2 MeV with a 176-μA time-averaged current hitting the braking target, assuming a beta relationship. Figure 3 depicts the photon flux created at the tungsten target. Integrating this flux over the energy range noted results in $1.7\times10^{14}$ photons per second per steradian. Over the range of these photon energies photoelectrons, Compton-scattered electrons, and pair-production electrons were created in the specimens. Though further work needs to be completed, it is believed there was some level of high-energy electrons that also made their way through the braking target. Glass-darkening experiments indicated a beam spread angle of about 45°. This considerable spread angle allowed the vials to be slightly displaced (~1.25 cm; 0.5 in.) from the beam centerline, allowing multiple tests to be performed simultaneously with axisymmetric near-peak flux and energy.

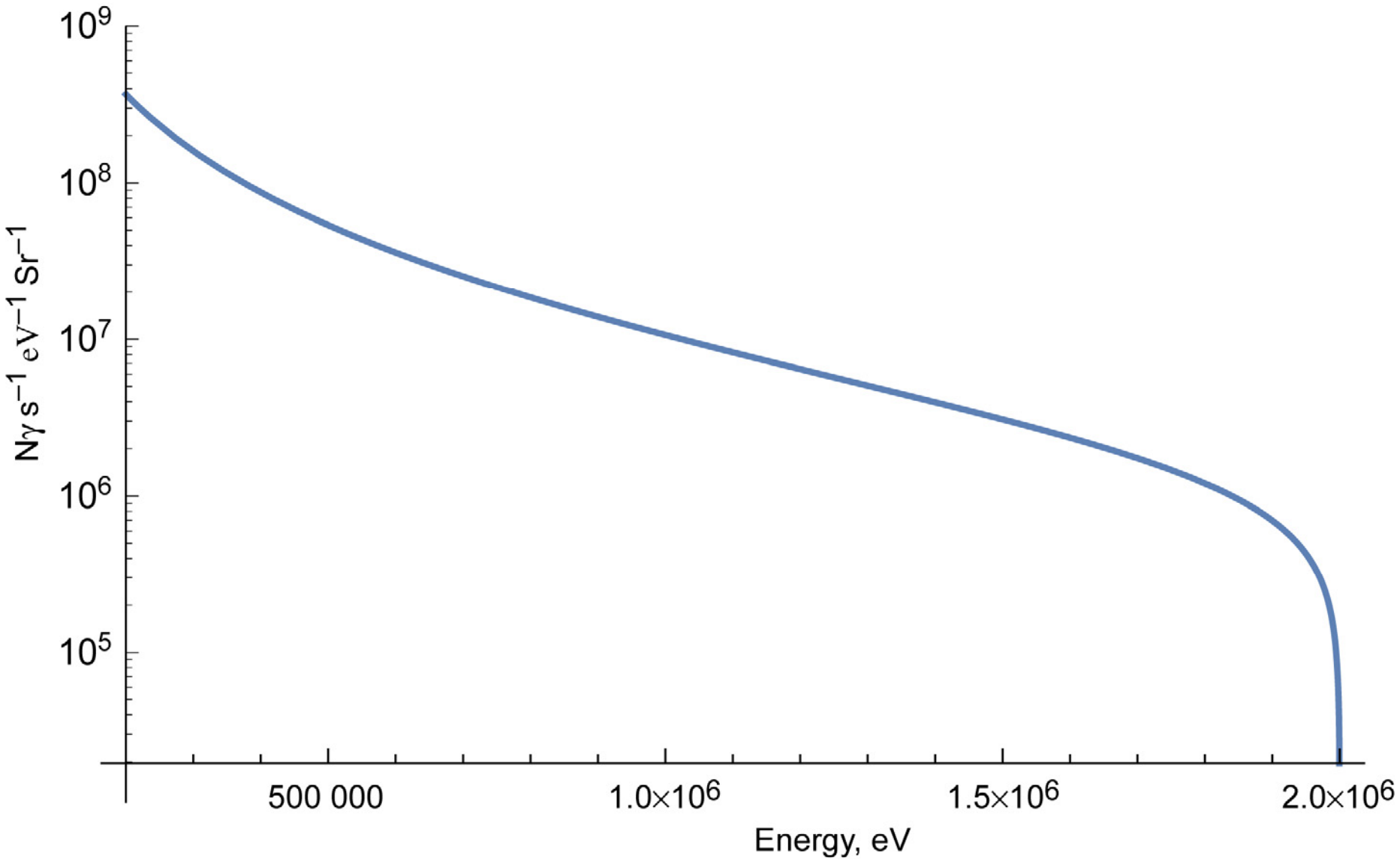

Figure 3.—Log of LINAC photon flux versus energy for the176-μA target current nominal 2-MeV energy setting (where $N_\gamma$ is number of photons).

### 2.2.2 End-Point Energy Measurement

Since an objective of this experiment was to explore possible reactions at photon energies below the deuteron photodissociation threshold, an important aspect of this study was to verify that the beam end-point energy was below the deuteron photodissociation threshold energy.

#### 2.2.2.1 Deuteron Photodissociation Energy

The threshold energy $E_{th}$ for photodissociation is found using the expression $E_{th} = BE(1 + BE/2m_d)$, where the binding energy ($BE$) is 2.225 MeV (Ref. 3) and the mass of the deuteron $m_d$ is 2.01410178$u$ (where $u$ is taken as 931.494 MeV per unit mass), yielding a value of the photodissociation energy of 2.226 MeV.

#### 2.2.2.2 Water Tank Ionization Chamber Method

Dosimetry-derived bremsstrahlung end-point energy was measured using an ion chamber and water table following Varian (LINAC manufacturer) practice consistent with Reference 9 (International Atomic Energy Agency, IAEA). In this method a calibrated ion gauge is placed on an x-y-z stage positioned within the water table. The system utilizes an automated program to spatially search for the maximum beam strength and once found, uses the measured value to determine the end-point energy. This method was repeated five times and obtained energies ranging from 1.93 to 1.975 MeV. Using the five measurements resulted in an average beam energy of 1.95 MeV. Based on the measurements, the beam end-point energy was determined to be between 1.78 and 2.13 MeV with a 4 sigma (99.99%) confidence level. Similar results were obtained by using multiple layers of steel plates. Consistent results indicated a nominal 1.95-MeV photon end-point energy of the same LINAC at both the vendor and repeated at NASA Plum Brook Station test facilities. It is realized that the dosimetry-measured dose does not directly measure the gamma end-point energy; however, the end-point energy is related to the ion gauge reading through a calibrated correlation equation. Both water table and steel plate results were consistent with the same nominal photon end-point energy. As a point of reference, because of the sealed beam construction, the team could not determine the peak electron acceleration voltage, a direct measure of the peak gamma end-point energy.

## 2.3 Nuclear Activity Measurements

Pre- and postexposure nuclear activity measurements were made using the instruments and techniques described in this section to evaluate the changes in specimen nuclear activity due to beam exposure.

### 2.3.1 Gamma Spectroscopy

To evaluate potential gamma activity and radioisotopes created by the LINAC bremsstrahlung exposure, materials used were scanned in their elemental components (prior to mixing) and scanned after bremsstrahlung exposure as a mixture. In every case the pre-exposed materials' spectral lines were indistinguishable from the natural background in the detector cave. After each exposure the contents of the 7-mL exposed vials were carefully decanted into new unexposed separate 20-mL glass vials to ensure that activity attributed only to the sample was measured. Note: For later tests with the streamlined process, the team gamma scanned all of the vials exposed at one time to capture the shorter half-lives.

Three separate gamma high-purity germanium (HPGe) detection systems were utilized to count the specimens. All were Ortec units employing cryocoolers, using cylindrical lead caves with passive graded shielding made of tin and copper to reach low background counts, generally less than 25 counts per second (cps). The cylindrical lead cave around the HPGe crystal has a 6.35-cm lid thickness, 6-cm wall thickness, 15.24-cm inner diameter, 27.3-cm outer diameter, and 21-cm chamber height. Units 1 (Mod. No. MX45P–A–S) and 2 (Mod. No. GMX45P4–A) utilized an aluminum window, allowing photon energies to be measured down to about 40 keV. Unit 3 (Mod GMX40P4) utilized a beryllium window, allowing x-ray lines to be measured down to 25 keV. Quality controls included daily spectral line checks with a certified check source containing $^{137}Cs$, $^{152}Eu$, and $^{241}Am$. The check source has a total activity of about 18.5 kBq (0.5 μCi) and known individual isotope activities as of the check source's issue date. Daily checks with the source include ensuring that the spectral lines of $^{137}Cs$, $^{152}Eu$, and $^{241}Am$ are at the proper position and that the detected activity (kBq) is as expected in accordance with the known activities of each isotope. The detector's minimal detectable activity is on the order of 5.55 Bq (0.15 nCi). Along with daily source checks, daily background checks with an empty, closed cave are performed to ensure that the lead cave is not contaminated. The daily check ensures the calibrated energy and efficiency for each detector is correct with an occasional small gain change as necessary to maintain specifications using check sources. NASA Glenn personnel perform periodic calibrations on the order of once per year or more often if needed. All units were within their calibration specifications (calibration date July 31, 2015). The manufacturer requires no calibration unless a detector cannot be brought into proper operating conditions. The spectra were displayed with GammaVision® 7 software (Ref. 10) using the NPP32 analysis engine. Gamma lines were checked also with the Lawrence Berkeley National Laboratory (LBNL) database (Ref. 11), amongst other sources, to confirm identity of the radioisotopes. For every test sample, at least a 15-min duration gamma scan was performed within an hour after beam-off to capture any short half-life radioisotopes. For the streamlined approach this was reduced to about a half hour. Also, a gamma scan of 1-h duration was performed on each sample after beam-off with the elapsed time as provided in Section 3.1. Follow-up scans were performed on several of the specimens to evaluate half-lives by examining the change in gamma activity levels. For short-half-life radioisotopes, multiple 15-min scans were completed. For longer half-life radioisotopes, longer scans, generally 1-h, were used. The team also used Cambio Software to aid in plotting and performing Gaussian area count statistics for some of the work performed herein (Ref. 12).

### 2.3.2 Beta Activity Measurement

Two instruments were used to measure beta activity. In each case, the materials used for the test were prescanned for alpha and beta activity using a Canberra (Tennelec) Series 5 XLB—Automatic Low Background Alpha/Beta Counting System gas flow proportional counter. Quality controls included daily checks using plated alpha ($^{239}$Pu) and beta ($^{99}$Tc) sources as well as an empty planchet to ensure proper system performance and acceptable background levels. This instrument was used to count the reactants before and after the exposure. In every case the pre-exposed materials' alpha and beta counts were indistinguishable from the background counts.

Beta scintillation measurements were done using a Beckman LS6500 beta scintillator for both baseline, or as-received, materials and a portion of postexposure materials. In cases where the counts were high, only a portion of the sample was used to keep the counts within the count limits of the instrument, and the other portion was kept separate for the Tennelec alpha/beta counting system scans described above. The LS6500 system was checked daily using a $^{14}$C check source according to Beckman specifications to confirm that the instrument was properly counting the corresponding standard. Environmentally friendly Ecoscint-O scintillator fluid was used. Whenever a sample was counted, a background count was performed for reference purposes. It was found during the test program that the postexposed D-paraffin exhibited high LUMEX levels (a measure of chemoluminescence), which limited the ability to measure beta energies.

### 2.3.3 Neutron Detection

Multiple instruments were used in the effort to measure neutron activity. The two electronic instruments utilized for the early tests were a Canberra Model AN/PDR–70 (Snoopy NP–2) neutron survey meter (with $BF_3$ media) and a Ludlum Model 2363 gamma-neutron survey meter (Prescilla). Both were shielded from the intense gamma energy using lead and positioned approximately 3.9 m from the source. However, it was discovered during testing that the electromagnetic interference (EMI) field created by the LINAC and magnetron was too strong, preventing reliable neutron measurements. The authors are investigating EMI shielding techniques for future tests.

Solid-state Bubble Detectors (Bubble Technology Industries Inc., model BD–PND) were also used to obtain relative measures of the neutron activity occurring during each experiment. Bubble Detector dosimeters are made of a polymer gel in which minute droplets of superheated liquid (e.g., Freon) are uniformly scattered (Ref. 13). When these droplets are struck by neutrons, they evaporate and form small bubbles of gas that remain trapped in the elastic polymer, providing a visible record of the activity. The number of bubbles is in direct proportion to the number of neutrons striking the detector. From the counted number of bubbles in the detectors and from their certified efficiency (bubbles/mREM), one can obtain the dose equivalent of the neutrons that struck the detector. These detectors proved useful because they are immune to gamma interference and EMI. Generally, two detectors were used to record the neutron activity during exposure. Each Bubble Detector was factory calibrated with a sensitivity ranging from 9 to 20 bubbles/mREM, which was noted in the logbooks. The advertised neutron energies for these detectors are quite broad: 100 to 200 keV up to 15 MeV. The Bubble Detectors were placed on top of the LINAC head approximately 17.8 cm (7 in.) away from where the test specimens were exposed. A temperature probe monitored the flange temperature, showing the detector temperature was maintained within the operating limits.

To quantify background neutron counts, several tests were performed. In some cases the beam was left OFF and in other cases the beam was turned ON, but with no fuel. In all background runs performed, either no bubbles or a single bubble was observed for the 4-h (beam OFF) and 6-h (beam ON) measurement intervals. Bubble counts of 1 to 2 are consistent with background in the earthen-covered concrete bunker housing the LINAC. This test provided evidence that the machine was not creating neutrons.

#### 2.3.3.1 Neutron Activation Materials

Finding that the electronic neutron detectors were not reliable, neutron activation materials were placed in separate vials placed either adjacent to the vials or on top of the LINAC head in a specially made cave of tungsten and lead. Different materials were used to help further refine the energies of the neutrons being observed by the Bubble Detectors. Table II gives the materials and neutron energies required to activate the stated channels (Ref. 14). Materials used included cadmium buttons (nom. 3 g), indium shavings (nom. 3 and 6 g), and gadolinium shavings (nom. 1 g). All of the materials were contained in 7-mL glass vials and gamma scanned prior to exposure to ensure no activity above background was present.

The cadmium and indium materials exhibited a metastable state when positioned in the LINAC beam. With the desire to eliminate this as the cause of the activity, a cave was built to shield the neutron activation materials on top of the LINAC. The cave (Fig. 1) used 10 cm (4 in.) of tungsten on the floor, front, ceiling, and back of the cave, and 5 cm (2 in.) of lead on the sides. The samples in the cave were approximately 30 cm (12 in.) from the braking target. This cave was shown to be ineffective during no-fuel test exposures in eliminating all gamma-induced metastable radioisotopes. Hence the $^{115m}$In data from this initial cave had to be discarded.

#### 2.3.3.2 Solid-State Nuclear Track Detectors

Solid-state nuclear track detectors (SSNTDs; i.e., CR–39, a polycarbonate resin), were also used to detect tracks of recoil protons caused by incident neutrons. CR–39's use as a SSNTD is described in Reference 15. When traversing a plastic material such as CR–39, charged particles such as recoil protons create a region along their ionization track that is more sensitive to chemical etching than the rest of the bulk (Ref. 16). The size, depth of penetration, and shape of these tracks provides information about the mass, charge, energy, and direction of motion of the particles (Ref. 17).

Rectangular-shaped detectors (1 cm × 2 cm × 1 mm, or 0.4 × 0.8 × 0.04 in.) made by Fukuvi USA, Inc., were obtained from Landauer, Inc. During each LINAC exposure, two CR–39 detectors were placed on top of the LINAC head about 18 cm (7 in.) from the braking target and out of the main gamma flux to record neutron and charged-particle strikes from the experiment. Two CR–39 detectors used as control detectors to record background charged-particle strikes were placed on a table on a lead brick (to simulate the lead LINAC head shielding) in the LINAC lab area but displaced ~11 m (~36 ft) from the LINAC beam and behind a ~1-m- (~39-in.-) thick concrete wall with lead shielding to shield the chips from the neutron field and prevent gamma damage. These were used to record background neutron levels in the laboratory for comparison to the exposed detectors. Two additional detectors were placed in the control room for the LINAC. Detectors were etched and then analyzed to establish neutron activity and neutron energies (Ref. 18).

TABLE II.—NEUTRON ACTIVATION MATERIALS, REACTION CHANNELS, AND ASSOCIATED NEUTRON ENERGIES TO CAUSE REACTION.[a]

| Sensor and material | Neutron energies | Reaction mechanism |
|---|---|---|
| Bubble Detectors[b] | 100 to 200 keV threshold through 15 MeV | Proprietary process: neutron interaction with superheated liquid |
| CR–39 polycarbonate ion track detectors | >144 keV, threshold 1 to 15 MeV, | Proton, carbon, and oxygen recoils For >10 MeV, $^{12}C_6$(n,n')3α (i.e., triple track) |
| Indium | Thermal to MeV level | $^{115}$In (n,γ) $^{116m}$In |
| | 0.336 MeV (threshold)[c] | $^{115}$In (n,n') $^{115m}$In |
| | 0.6 MeV (10 mb)[c] | $^{115}$In (n,n') $^{115m}$In |
| | 1.2 MeV (0.5 b)[c] | $^{115}$In (n,n') $^{115m}$In |
| | 1.7 MeV (1 b)[c] | $^{115}$In (n,n') $^{115m}$In |
| Gadolinium | Thermal to MeV level | $^{158}$Gd (n,γ) $^{159}$Gd |
| Cadmium | Thermal to MeV level | $^{114}$Cd (n,γ) $^{115}$Cd |

[a]Only channels with adequate half-lives that could be detected after exposure were included.
[b]Bubble Technology Industries Inc.
[c]Neutron capture cross sections obtained from Janis Book (Ref. 19).

# 3.0 Experimental Results

## 3.1 Gamma Spectroscopy

Gamma spectroscopy was used to measure nuclear activity for specimens with and without deuterium after exposure and to assist in half-life determination.

### 3.1.1 Deuterated Materials

Each as-received material was scanned for gamma spectral lines before beam exposure, and none was found to have any activity different than the standard background lines. During the test phase, it took approximately 45 to 60 min after beam exposure to move the sample from the LINAC test head, get it ready for scanning, and place it into the HPGe detector. In preparation for scanning, the sample was decanted from the exposed 7-ml vial, placed into a nonexposed 20-ml vial, and counted for 15 min, and then it underwent a 60-min HPGe count. The deuterated erbium and hafnium samples all showed gamma activity after exposure.

Peaks significantly above background from the collected gamma spectra were analyzed and separated into groups corresponding to the following radioisotopes based on the material tested: $^{163}$Er/$^{165}$Er, $^{171}$Er, $^{180m}$Hf, $^{181}$Hf, $^{99}$Mo, $^{101}$Mo, $^{99m}$Tc, and $^{101}$Tc. Radioisotope identification was confirmed by identifying that each radioisotope's strongest gamma line was present, confirming that subsequent gamma lines in accordance with their known levels of intensity were present, and calculated half-lives (based on measurements) of each peak coincided with their respective known radioisotope half-life. Two peaks from 2 naturally occurring radioactive daughters of radon ($^{222}$Rn), $^{210}$Pb, and $^{214}$Pb are located near peaks found in the collected gamma spectra at 46.70 keV (attributed to $^{163}$Er/$^{165}$Er) and 295.94 keV (attributed to $^{171}$Er). Two additional peaks at 111.62 keV (attributed to $^{171}$Er) and 443.09 keV (attributed to $^{180m}$Hf) were found that could belong to 2 other naturally occurring radioisotopes of $^{234}$Th and $^{223}$Ra respectively. However, the strongest gamma line of $^{234}$Th (63.29 keV) was not present in gamma spectra collected with samples containing deuterated erbium. In addition, the strongest gamma line of $^{223}$Ra (269.46 keV) was not present in any of the gamma spectra collected. The calculated half-lives of all 4 peaks did not coincide with the half-lives of $^{210}$Pb, $^{234}$Th, $^{214}$Pb, and $^{223}$Ra. Instead, the half-lives were much closer to those of $^{163}$Er/$^{165}$Er, $^{171}$Er, and $^{180m}$Hf. Table III shows the gamma lines, gamma line intensity, and half-lives for each of the seven aforementioned radioisotopes. Although counts from these naturally occurring radioisotopes could be within the four peaks observed in the gamma spectra, the peaks (with reasons stated above) are attributed to the radioisotopes created within the tested materials, as will be further described below.

Tables IV(a) and V(a) show for these $ErD_{2.8}$ and $HfD_2$ samples, respectively, the new radioisotopes found that correspond to the accepted radioisotope energies. Tables IV(b) and V(b) show gamma results for additional $ErD_x$ and $HfD_2$ samples, respectively, that were transported back to the HPGE detector using a streamlined technique allowing faster entry (~30 min) into the HPGe detector and measurement of shorter lived radioisotopes. The tables include the corresponding net area counts (significantly above background counts), uncertainty (1 sigma), and full width half maximum energy used for analysis. These results were computed using the GammaVision® 7 software protocol (Ref. 10).

For the deuterated erbium specimens, Table IV(a) shows posttest gamma spectra, identifying evidence of radioisotopes of erbium ($^{163}$Er and $^{171}$Er) and molybdenum ($^{99}$Mo) and via beta decay, technetium ($^{99m}$Tc). For the deuterated erbium specimens, Table IV(b) shows posttest gamma spectra, identifying evidence of radioisotopes of erbium ($^{163}$Er and $^{171}$Er) and molybdenum ($^{99}$Mo and $^{101}$Mo) and via beta decay, technetium ($^{99m}$Tc and $^{101}$Tc).

Table V(a) shows posttest gamma spectra, identifying evidence of radioisotopes of erbium ($^{180m}$Hf and $^{181}$Hf) and molybdenum ($^{99}$Mo) and via beta decay, technetium ($^{99m}$Tc) for the deuterated hafnium specimens. Table V(b) shows posttest gamma spectra identifying evidence of radioisotopes of erbium ($^{180m}$Hf and $^{181}$Hf) and molybdenum ($^{99}$Mo and $^{101}$Mo) and via beta decay, technetium ($^{99m}$Tc and $^{101}$Tc).

TABLE III.—COMPARISON OF MEASURED PEAK GAMMA LINES WITH NATURALLY OCCURING BACKGROUND RADIOISOTOPES AND ERBIUM OR HAFNIUM RADIOISOTOPES
[Accepted gamma energy data from Lawrence Berkeley National Laboratory (LBNL, Ref. 11).]

| Experimental data | | Accepted data per LBNL<br>Radioisotope gamma lines[a] | | | |
|---|---|---|---|---|---|
| Specimen | Measured peak/<br>Calculated half-life ($t_{1/2}$) | Naturally occurring background | | Created from test material | |
| PGL2150-2153 | 46.47 keV/42.67 min | **$^{210}$Pb ($t_{1/2}$ = 22.3 y)** | | **$^{163}$Er: x-ray lines ($t_{1/2}$ = 75.0 min)** | |
| | | Energy, keV | Intensity, percent | Energy, keV | Intensity, percent |
| | | *46.54* | *4.25* | 47.55 | 39.90 |
| | | | | *46.70* | *22.40* |
| | | | | 53.88 | 7.98 |
| | | | | 6.72 | 6.70 |
| PGL2150-2153 | 111.57 keV/16.46 h | **$^{234}$Th ($t_{1/2}$ = 24.1 days)** | | **$^{171}$Er ($t_{1/2}$ = 7.52 h)** | |
| | | Energy, keV | Intensity, percent | Energy, keV | Intensity, percent |
| | | 63.29 | 4.80 | 308.31 | 64.40 |
| | | 92.38 | 2.81 | 295.90 | 28.90 |
| | | 92.80 | 2.77 | 50.74 | 23.50 |
| | | *112.81* | *0.28* | *111.62* | *20.50* |
| PGL2150-2153 | 295.88 keV/4.79 h | **$^{214}$Pb ($t_{1/2}$ = 26.8 min)** | | **$^{171}$Er ($t_{1/2}$ = 7.52 h)** | |
| | | Energy, keV | Intensity, percent | Energy, keV | Intensity, percent |
| | | 351.93 | 37.60 | 308.31 | 64.40 |
| | | *295.22* | *19.30* | *295.90* | *28.90* |
| | | 241.99 | 7.43 | 50.74 (x-ray) | 23.50 |
| | | 53.23 | 1.20 | 111.62 | 20.50 |
| PGL2142-2145 | 443.41 keV/11.21 h | **$^{223}$Ra ($t_{1/2}$ = 11.44 days)** | | **$^{180m}$Hf ($t_{1/2}$ = 5.5 h)** | |
| | | Energy, keV | Intensity, percent | Energy, keV | Intensity, percent |
| | | 269.46 | 13.70 | 332.28 | 94.10 |
| | | 154.21 | 5.62 | *443.09* | *81.90* |
| | | 323.87 | 3.93 | 215.26 | 81.30 |
| | | 144.23 | 3.22 | 57.56 | 48.00 |
| | | 338.28 | 2.79 | 93.33 | 17.10 |
| | | *445.03* | *1.27* | 500.64 | 14.30 |

[a]Italic numbers were those values that were compared.

TABLE IV.—$ErD_{2.8}+C_{36}D_{74}$+Mo GAMMA LINES
[Accepted gamma energy data from Lawrence Berkeley National Laboratory (LBNL, Ref. 11).]

(a) PGL 2028a, 2029aa, and 2050 to 2053, measured after 4-h exposure (60-min scan time)

| Experiment details | | Postexposure new radioisotopes identified | Accepted data per LBNL | | Experimental data | | | |
|---|---|---|---|---|---|---|---|---|
| Specimen | Elapsed time from end of shot (detector number) | | Gamma line energy, keV | Intensity, percent | Centroid energy, keV | Net area counts | Count uncertainty, ± | Full width half maximum (FWHM) |
| PGL 2028a | 6.75 h (Unit 1) | $^{163}$Er | 47.55 | 40 | 47.23 | 1923 | 166 | 1.48 |
| | | $^{171}$Er | 111.62 | 20 | 111.65 | 675 | 38 | 0.86 |
| | | $^{171}$Er | 116.66 | 2 | 116.58 | 43 | 28 | 0.59 |
| | | $^{171}$Er | 295.90 | 29 | 296.01 | 1353 | 43 | 1.08 |
| | | $^{171}$Er | 308.31 | 64 | 308.41 | 2916 | 58 | 1.07 |
| | | $^{99}$Mo | 140.51 | 89 | 140.6 | 1068 | 62 | 0.93 |
| | | $^{99}$Mo | 181.06 | 6 | 181.1 | 43 | 27 | 1.1 |
| | | $^{99}$Mo | 739.50 | 12 | 739.33 | 98 | 17 | 1.19 |
| | | $^{99}$Mo | 777.92 | 4 | 778.23 | 48 | 15 | 0.37 |
| PGL 2029aa | 6.75 h (Unit 3) | $^{163}$Er | 47.55 | 40 | 47.52 | 1322 | 104 | 2.05 |
| | | $^{171}$Er | 111.62 | 20 | 111.67 | 504 | 32 | 1.24 |
| | | $^{171}$Er | 116.66 | 2 | 117.11 | 60 | 28 | 0.36 |
| | | $^{171}$Er | 124.02 | 9 | 124.1 | 242 | 31 | 0.91 |
| | | $^{171}$Er | 295.90 | 29 | 296.06 | 1004 | 40 | 1.18 |
| | | $^{171}$Er | 308.31 | 64 | 308.48 | 2101 | 60 | 1.26 |
| | | $^{99}$Mo | 140.51 | 89 | 140.56 | 698 | 38 | 1.02 |
| | | $^{99}$Mo | 181.06 | 6 | 180.83 | 76 | 53 | 0.44 |
| | | $^{99}$Mo | 739.50 | 12 | 740.06 | 76 | 16 | 0.8 |
| | | $^{99}$Mo | 777.92 | 4 | 777.95 | 32 | 15 | 0.54 |
| PGL 2050 | 3.25 h (Unit 1) | $^{163}$Er | 47.55 | 40 | 47.19 | 2398 | 154 | 1.88 |
| | | $^{171}$Er | 111.62 | 20 | 111.61 | 1182 | 45 | 0.88 |
| | | $^{171}$Er | 116.66 | 2 | 116.59 | 108 | 33 | 0.59 |
| | | $^{171}$Er | 124.02 | 9 | 124 | 567 | 39 | 0.86 |
| | | $^{171}$Er | 295.90 | 29 | 295.95 | 2516 | 55 | 1.01 |
| | | $^{171}$Er | 308.31 | 64 | 308.34 | 5306 | 75 | 1.08 |
| | | $^{99}$Mo | 140.51 | 89 | 140.52 | 1072 | 80 | 1.07 |
| | | $^{99}$Mo | 181.06 | 6 | 181.2 | 127 | 28 | 0.47 |
| | | $^{99}$Mo | 739.50 | 12 | 739.28 | 133 | 20 | 1.22 |
| | | $^{99}$Mo | 777.92 | 4 | 777.59 | 78 | 15 | 1.07 |
| PGL 2051 | 3.5 h (Unit 3) | $^{163}$Er | 47.55 | 40 | 47.63 | 3021 | 154 | 1.92 |
| | | $^{171}$Er | 111.62 | 20 | 111.67 | 1484 | 48 | 1.28 |
| | | $^{171}$Er | 116.66 | 2 | 116.77 | 247 | 36 | 0.96 |
| | | $^{171}$Er | 124.02 | 9 | 124.06 | 738 | 44 | 1.01 |
| | | $^{171}$Er | 295.90 | 29 | 295.95 | 3314 | 62 | 1.28 |
| | | $^{171}$Er | 308.31 | 64 | 308.39 | 6777 | 85 | 1.27 |
| | | $^{99}$Mo | 140.51 | 89 | 140.51 | 1270 | 52 | 1.07 |
| | | $^{99}$Mo | 181.06 | 6 | 181.09 | 231 | 64 | 1.19 |
| | | $^{99}$Mo | 739.50 | 12 | 739.85 | 129 | 22 | 1.77 |
| | | $^{99}$Mo | 777.92 | 4 | 778.11 | 85 | 18 | 1.77 |
| PGL 2052 | 4.5 h (Unit 3) | $^{163}$Er | 47.55 | 40 | 47.51 | 2933 | 166 | 1.64 |
| | | $^{171}$Er | 111.62 | 20 | 111.66 | 1539 | 50 | 1.07 |
| | | $^{171}$Er | 116.66 | 2 | 116.72 | 193 | 36 | 0.98 |
| | | $^{171}$Er | 124.02 | 9 | 124.05 | 850 | 45 | 1.27 |
| | | $^{171}$Er | 295.90 | 29 | 295.99 | 3327 | 63 | 1.23 |
| | | $^{171}$Er | 308.31 | 64 | 308.41 | 6891 | 86 | 1.32 |
| | | $^{99}$Mo | 140.51 | 89 | 140.56 | 1772 | 57 | 1.12 |
| | | $^{99}$Mo | 181.06 | 6 | 181.15 | 175 | 32 | 1.09 |
| | | $^{99}$Mo | 739.50 | 12 | 739.94 | 216 | 19 | 1.22 |
| | | $^{99}$Mo | 777.92 | 4 | 778.35 | 59 | 18 | 0.74 |

TABLE IV.—$ErD_{2.8}$+$C_{36}D_{74}$+Mo GAMMA LINES (Concluded)
[Accepted gamma energy data from Lawrence Berkeley National Laboratory (LBNL, Ref. 11).]

(a) PGL 2028a, 2029aa, and 2050 to 2053, measured after 4-h exposure (60-min scan time)

| Experiment details | | Postexposure new radioisotopes identified | Accepted data per LBNL | | Experimental data | | | |
|---|---|---|---|---|---|---|---|---|
| Specimen | Elapsed time from end of shot (detector number) | | Gamma line energy, keV | Intensity, percent | Centroid energy, keV | Net area counts | Count uncertainty, ± | Full width half maximum (FWHM) |
| PGL 2053 | 4.5 h (Unit 1) | $^{163}Er$ | 47.55 | 40 | 47.37 | 3676 | 163 | 1.85 |
| | | $^{171}Er$ | 111.62 | 20 | 111.65 | 1611 | 51 | 0.85 |
| | | $^{171}Er$ | 116.66 | 2 | 116.72 | 243 | 36 | 1.08 |
| | | $^{171}Er$ | 124.02 | 9 | 124.08 | 667 | 106 | 0.97 |
| | | $^{171}Er$ | 295.90 | 29 | 295.96 | 3489 | 78 | 1.11 |
| | | $^{171}Er$ | 308.31 | 64 | 308.36 | 7295 | 88 | 1.11 |
| | | $^{99}Mo$ | 140.51 | 89 | 140.57 | 1547 | 128 | 1.01 |
| | | $^{99}Mo$ | 181.06 | 6 | 181.24 | 161 | 32 | 0.74 |
| | | $^{99}Mo$ | 739.50 | 12 | 739.34 | 212 | 30 | 1.33 |
| | | $^{99}Mo$ | 777.92 | 4 | 777.99 | 89 | 16 | 0.72 |
| (b) PGL 2150 TO 2153 measured after 6-h exposure; streamlined transport from LINAC to HPGe (15-min scan time) | | | | | | | | |
| PGL 2150 to 2153 | 0.55 h Unit 1 | $^{163}Er$ | 47.55 | 40 | 47.37 | 1355 | 105 | 1.35 |
| | | $^{171}Er$ | 111.62 | 20 | 111.56 | 537 | 31 | 0.93 |
| | | $^{171}Er$ | 116.66 | 2 | Peak not present | | | |
| | | $^{171}Er$ | 124.02 | 9 | 124.04 | 314 | 49 | 0.92 |
| | | $^{171}Er$ | 295.90 | 29 | 295.88 | 1478 | 42 | 1.13 |
| | | $^{171}Er$ | 308.31 | 64 | 308.08 | 4443 | 105 | 1.973 |
| | | $^{101}Tc$ | 306.86 | 89 | Too close to $^{171}Er$ line at 308.31 | | | |
| | | $^{101}Tc$ | 545.12 | 6 | 544.92 | 71 | 12 | 1.13 |
| | | $^{101}Mo$ | 590.10 | 19 | 590.57 | 95 | 12 | 0.98 |
| | | $^{101}Mo$ | 191.92 | 18 | 192.03 | 138 | 22 | 0.68 |
| | | $^{101}Mo$ | 1012.47 | 13 | 1012.04 | 47 | 9 | 2.15 |
| | | $^{101}Mo$ | 505.92 | 12 | 505.76 | 32 | 12 | 0.78 |
| | | $^{99m}Tc$ | 140.51 | 89 | 140.51 | 216 | 94 | 0.89 |
| | | $^{99}Mo$ | 140.51 | 89 | 140.51 | 216 | 94 | 0.89 |
| | | $^{99}Mo$ | 181.06 | 6 | 181.09 | 13 | 54 | 0.79 |
| | | $^{99}Mo$ | 739.50 | 12 | 739.4 | 69 | 12 | 1.34 |
| | | $^{99}Mo$ | 777.92 | 4 | 777.6 | 20 | 12 | 1.42 |

TABLE V.—$HfD_2$+$C_{36}D_{74}$+Mo GAMMA LINES

[Accepted gamma energy data from Lawrence Berkeley National Laboratory (LBNL, Ref. 11).]

(a) PGL 2034, 2035, and 2068 to 2071, measured after 4-h exposure (60-min scan time)

| Experiment details | | Closest radioisotope based on published data | Accepted data per LBNL | | Experimental data | | | |
|---|---|---|---|---|---|---|---|---|
| Specimen | Elapsed time from end of shot (detector number) | | Accepted gamma line energy, keV | Intensity, percent | Centroid energy, keV | Net area counts | Count uncertainty, ± | Full width half maximum (FWHM) |
| PGL 2034 | 4.5 h (Unit 1) | $^{180m}Hf$ | 215.26 | 81 | 215.43 | 375 | 28 | 1.09 |
| | | $^{180m}Hf$ | 332.28 | 94 | 332.37 | 440 | 37 | 1.09 |
| | | $^{180m}Hf$ | 443.09 | 82 | 443.22 | 282 | 38 | 0.91 |
| | | $^{181}Hf$ | 133.02 | 43 | Peak present, but not significantly above background | | | |
| | | $^{181}Hf$ | 345.92 | 15 | 345.81 | 30 | 19 | 0.49 |
| | | $^{181}Hf$ | 482.18 | 80 | 482.17 | 140 | 23 | 1.32 |
| | | $^{99}Mo$ | 140.51 | 89 | 140.58 | 1383 | 112 | 0.87 |
| | | $^{99}Mo$ | 181.06 | 6 | 181.21 | 186 | 50 | 1 |
| | | $^{99}Mo$ | 739.50 | 12 | 739.8 | 96 | 19 | 0.97 |
| | | $^{99}Mo$ | 777.92 | 4 | 778.14 | 38 | 19 | 0.75 |
| PGL 2035 | 4.5 h (Unit 3) | $^{180m}Hf$ | 215.26 | 81 | 215.28 | 354 | 28 | 1.29 |
| | | $^{180m}Hf$ | 332.28 | 94 | 332.39 | 357 | 38 | 1.3 |
| | | $^{180m}Hf$ | 443.09 | 82 | 443.4 | 235 | 22 | 1.12 |
| | | $^{180m}Hf$ | 500.64 | 14 | 501.04 | 60 | 18 | 0.5 |
| | | $^{181}Hf$ | 133.02 | 43 | 133.13 | 108 | 44 | 1.1 |
| | | $^{181}Hf$ | 345.92 | 15 | 346.11 | 36 | 20 | 0.62 |
| | | $^{181}Hf$ | 482.18 | 80 | 482.72 | 141 | 20 | 1.03 |
| | | $^{99}Mo$ | 140.51 | 89 | 140.55 | 918 | 41 | 1.12 |
| | | $^{99}Mo$ | 181.06 | 6 | 181.02 | 95 | 27 | 0.86 |
| | | $^{99}Mo$ | 739.50 | 12 | 740.05 | 78 | 21 | 1.04 |
| | | $^{99}Mo$ | 777.92 | 4 | 778.48 | 47 | 18 | 1.55 |
| PGL 2068 | 4.5 h (Unit 1) | $^{180m}Hf$ | 215.26 | 81 | 215.34 | 226 | 27 | 0.81 |
| | | $^{180m}Hf$ | 332.28 | 94 | 332.4 | 331 | 25 | 1.29 |
| | | $^{180m}Hf$ | 443.09 | 82 | 443.04 | 254 | 34 | 1.2 |
| | | $^{181}Hf$ | 133.02 | 43 | 132.99 | 59 | 26 | 0.87 |
| | | $^{181}Hf$ | 482.18 | 15 | 481.97 | 96 | 21 | 1.29 |
| | | $^{99}Mo$ | 140.51 | 89 | 140.54 | 998 | 39 | 0.86 |
| | | $^{99}Mo$ | 181.06 | 6 | 181.25 | 130 | 24 | 0.81 |
| | | $^{99}Mo$ | 739.50 | 12 | 739.32 | 121 | 18 | 1.43 |
| | | $^{99}Mo$ | 777.92 | 4 | 777.83 | 21 | 18 | 0.44 |
| PGL 2069 | 3.5 h (Unit 3) | $^{180m}Hf$ | 215.26 | 81 | 215.18 | 463 | 29 | 1.18 |
| | | $^{180m}Hf$ | 332.28 | 94 | 332.43 | 409 | 27 | 1.19 |
| | | $^{180m}Hf$ | 443.09 | 82 | 443.32 | 301 | 25 | 1.45 |
| | | $^{181}Hf$ | 133.02 | 43 | Peak present but not significantly above background | | | |
| | | $^{181}Hf$ | 345.92 | 15 | 345.86 | 25 | 19 | 0.32 |
| | | $^{181}Hf$ | 482.18 | 80 | 482.21 | 104 | 31 | 1.3 |
| | | $^{99}Mo$ | 140.51 | 89 | 140.49 | 1104 | 99 | 1.18 |
| | | $^{99}Mo$ | 181.06 | 6 | 180.93 | 285 | 50 | 1.11 |
| | | $^{99}Mo$ | 739.50 | 12 | 739.79 | 117 | 20 | 1.03 |
| | | $^{99}Mo$ | 777.92 | 4 | 778.45 | 37 | 18 | 0.62 |
| PGL 2070 | 4.5 h (Unit 3) | $^{180m}Hf$ | 215.26 | 81 | 215.19 | 269 | 28 | 1.15 |
| | | $^{180m}Hf$ | 332.28 | 94 | 332.42 | 300 | 26 | 1.29 |
| | | $^{180m}Hf$ | 443.09 | 82 | 443.41 | 210 | 34 | 0.98 |
| | | $^{181}Hf$ | 133.02 | 43 | Peak present but not significantly above background | | | |
| | | $^{181}Hf$ | 345.92 | 15 | Peak present but not significantly above background | | | |
| | | $^{181}Hf$ | 482.18 | 80 | 482.57 | 100 | 45 | 0.71 |
| | | $^{99}Mo$ | 140.51 | 89 | 140.52 | 908 | 81 | 1.21 |
| | | $^{99}Mo$ | 181.06 | 6 | 181.09 | 103 | 56 | 0.39 |
| | | $^{99}Mo$ | 739.50 | 12 | 739.83 | 121 | 19 | 0.83 |
| | | $^{99}Mo$ | 777.92 | 4 | 777.32 | 59 | 16 | 2.37 |
| PGL 2071 | 6 h (Unit 1) | $^{180m}Hf$ | 215.26 | 81 | 215.29 | 223 | 25 | 1.09 |
| | | $^{180m}Hf$ | 332.28 | 94 | 332.27 | 209 | 24 | 1.02 |
| | | $^{180m}Hf$ | 443.09 | 82 | 443.02 | 179 | 22 | 1.37 |
| | | $^{181}Hf$ | 133.02 | 43 | Peak present but not significantly above background | | | |
| | | $^{181}Hf$ | 345.92 | 15 | Peak present but not significantly above background | | | |
| | | $^{181}Hf$ | 482.18 | 80 | 482.2 | 70 | 20 | 1.24 |
| | | $^{99}Mo$ | 140.51 | 89 | 140.53 | 869 | 38 | 0.9 |
| | | $^{99}Mo$ | 739.50 | 12 | 739.11 | 129 | 17 | 1.18 |
| | | $^{99}Mo$ | 777.92 | 4 | 778.1 | 29 | 18 | 0.4 |

TABLE V.—$HfD_2+C_{36}D_{74}$+Mo GAMMA LINES (Concluded)

[Accepted gamma energy data from Lawrence Berkeley National Laboratory (LBNL, Ref. 11).]

(b) PGL 2142 to 2145 gamma lines measured after 6-h exposure; streamlined transport from LINAC TO HPGe (15-min scan time)

| Experiment details | | Closest radioisotope based on published data | Accepted data per LBNL | | Experimental data | | | |
|---|---|---|---|---|---|---|---|---|
| Specimen | Elapsed time from end of shot (detector number) | | Accepted gamma line energy, keV | Intensity, percent | Centroid energy, keV | Net area counts | Count uncertainty, ± | Full width half maximum (FWHM) |
| PGL 2142 to 2145 | 0.567 h Unit 3 | $^{180m}Hf$ | 215.26 | 81 | 215.3 | 359 | 29 | 1.1 |
| | | $^{180m}Hf$ | 332.28 | 94 | 332.49 | 417 | 27 | 0.91 |
| | | $^{180m}Hf$ | 443.09 | 82 | 443.43 | 309 | 23 | 1.22 |
| | | $^{181}Hf$ | 133.02 | 43 | 133.23 | [a]71 | 42 | 0.9 |
| | | $^{181}Hf$ | 345.92 | 15 | 346.18 | 41 | 17 | 0.62 |
| | | $^{181}Hf$ | 482.18 | 80 | 482.54 | 117 | 18 | 1.23 |
| | | $^{101}Tc$ | 306.86 | 89 | 307.02 | 4825 | 78 | 1.26 |
| | | $^{101}Tc$ | 545.12 | 6 | 545.58 | 184 | 22 | 1.12 |
| | | $^{101}Mo$ | 590.10 | 19 | 591.35 | [a]57 | 20 | 1.21 |
| | | $^{101}Mo$ | 191.92 | 18 | 191.99 | 332 | 47 | 1.21 |
| | | $^{101}Mo$ | 1012.47 | 13 | 1013.32 | 121 | 14 | 1.34 |
| | | $^{101}Mo$ | 505.92 | 12 | 506.19 | 168 | 34 | 1.11 |
| | | $^{99m}Tc$ | 140.51 | 89 | 140.62 | 554 | 53 | 1.19 |
| | | $^{99}Mo$ | 140.51 | 89 | 140.62 | 554 | 53 | 1.19 |
| | | $^{99}Mo$ | 181.06 | 6 | 181.22 | 252 | 80 | 1.31 |
| | | $^{99}Mo$ | 739.50 | 12 | 739.92 | 123 | 21 | 1.34 |
| | | $^{99}Mo$ | 777.92 | 4 | 778.29 | 53 | 13 | 0.46 |

[a]Net area obtained by scaling 60-min scan results.

Figure 4 shows example overall gamma spectra for samples PGL 2150 to 2153, all four vials counted for 15 min. The 308.3 keV gamma line corresponding to $^{171}Er$ showed the greatest activation at more than 2650 counts/keV (>30× background). Other characteristic lines are annotated for reference purposes.

Figure 5 shows example overall gamma spectra for PGL 2142 to 2145, counted for 15 min. The 306.9-keV line corresponding to $^{101}Tc$ had the highest number activation at 3612 counts/keV (>100 × background). Other characteristic lines are annotated for reference purposes.

It can be seen from both sets of data the photon stimulation technique with deuterated materials created new isotopes with 100 percent reproducibility. It is interesting to note, comparing Figures 4 and 5, that the $^{101}Tc$ line observed with the $HfD_2$/D-para/Mo sample was greater than twice the counts of the same line observed with the $ErD_{2.8}$/D-para/Mo. The deuterated erbium and hafnium material systems resulted in comparable activation of the gamma lines (140.5 keV) shared by $^{99}Mo$ and $^{99m}Tc$.

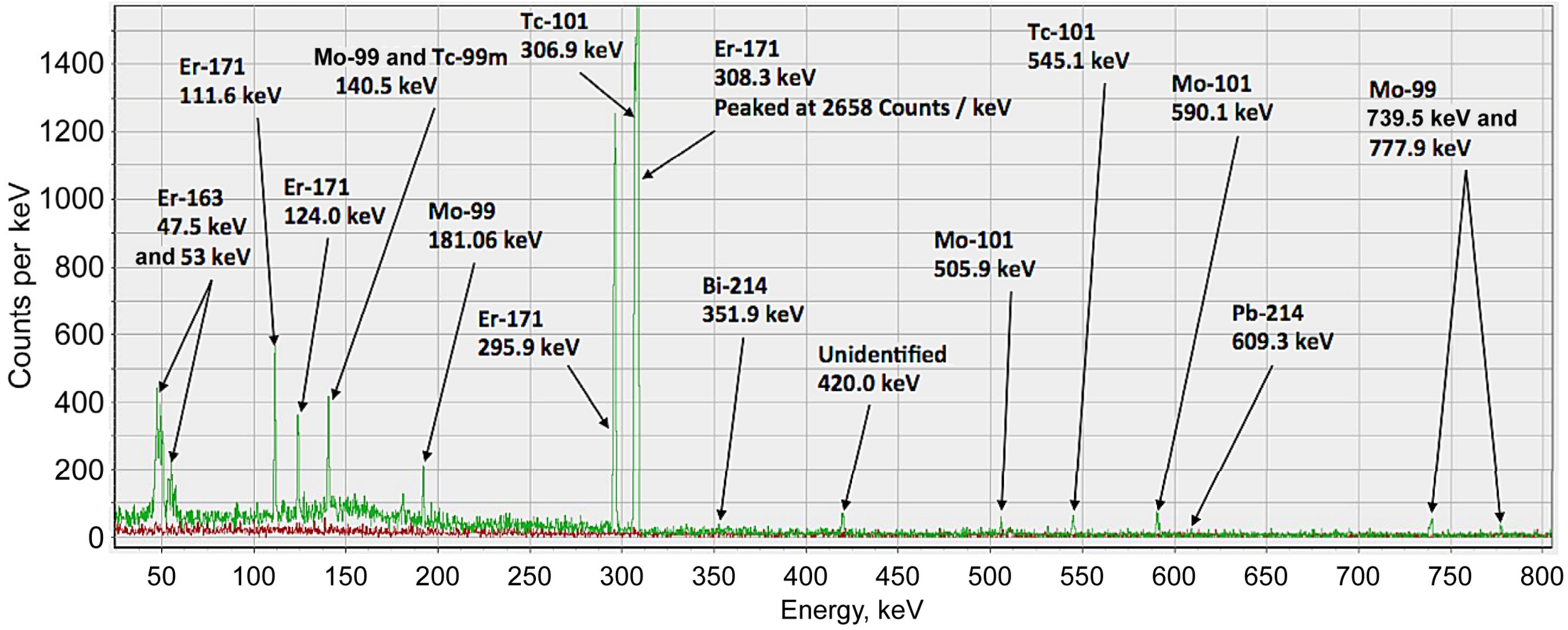

Figure 4.—Sample PGL 2150 to 2153: $ErD_{2.8}+C_{36}D_{74}$+Mo gamma spectra after 6-h exposure; 15-min counting interval. Green: sample results; red: cave background.

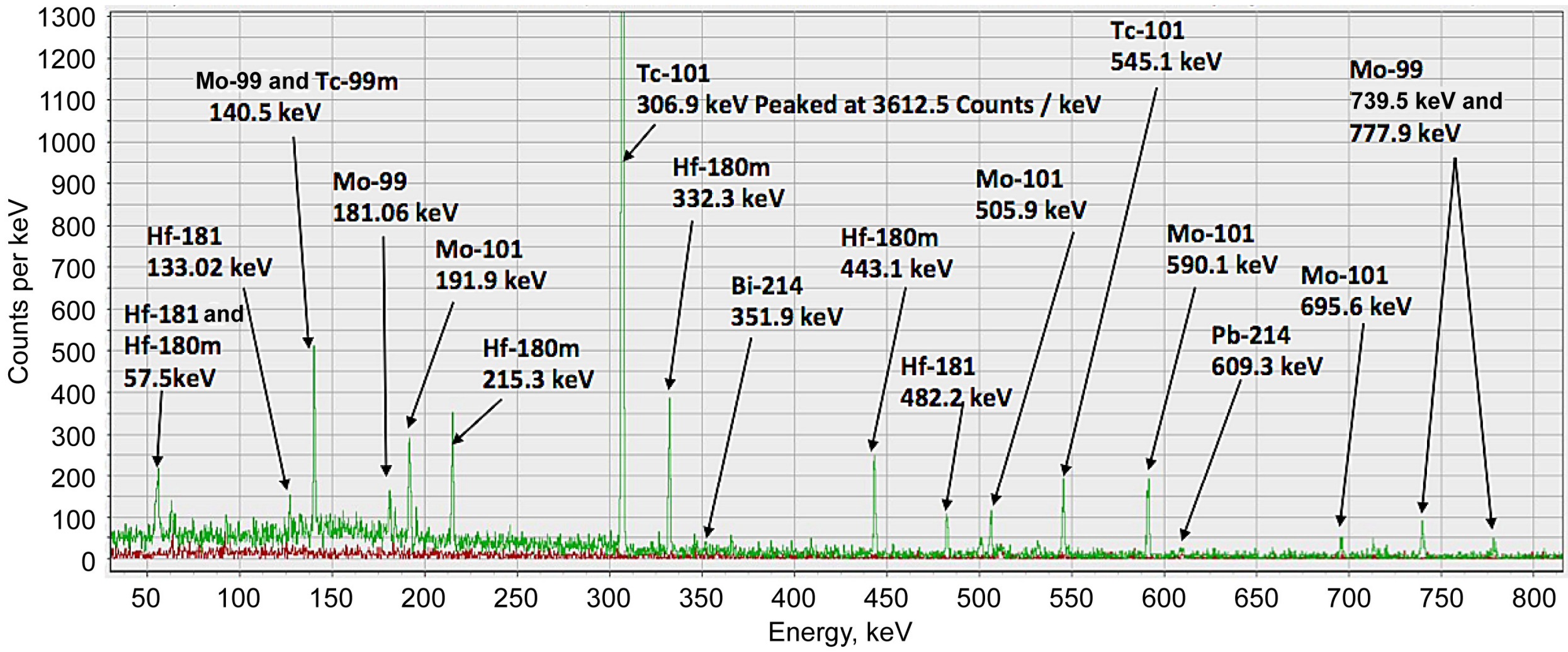

Figure 5.—Sample PGL 2142 to 2145: $HfD_2+C_{36}D_{74}$+Mo gamma spectra after 6-h exposure; 15-min counting interval. Green: sample results; red: cave background.

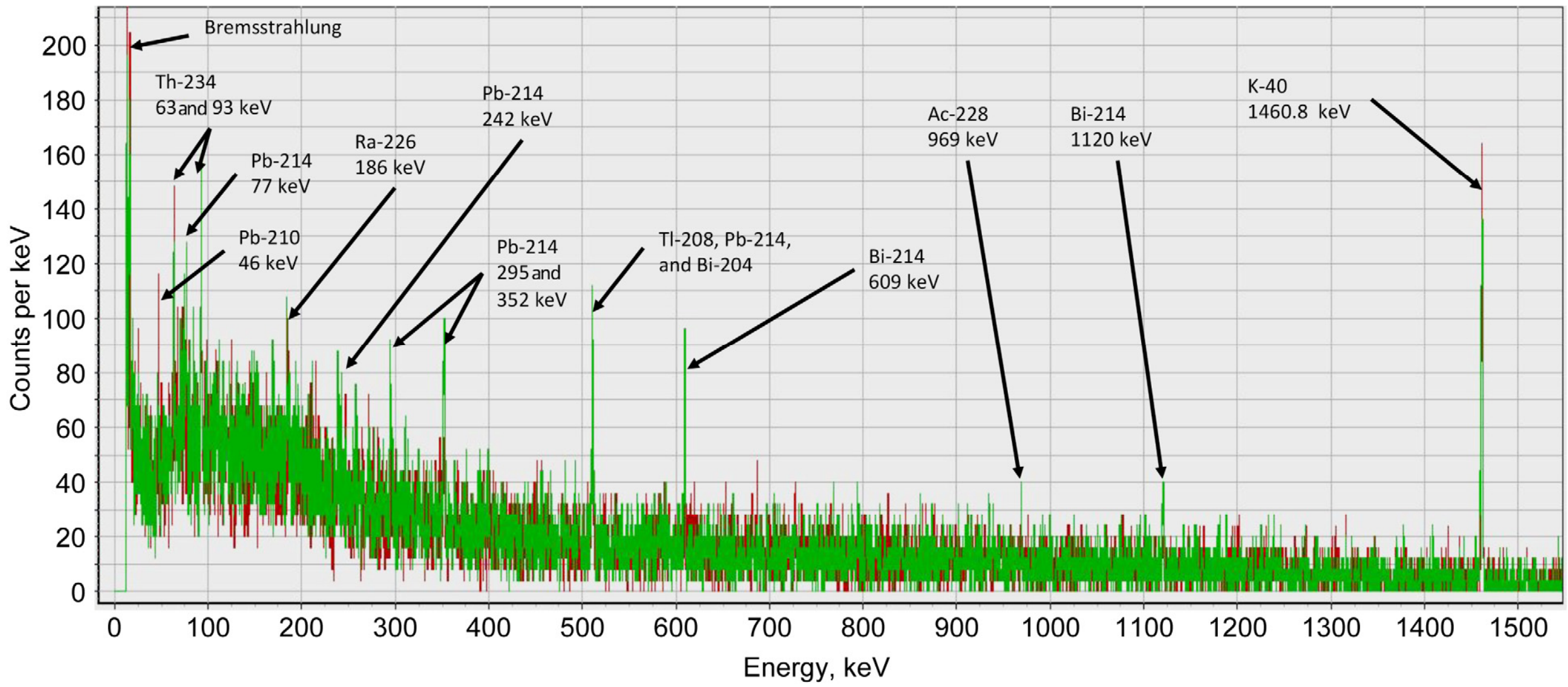

Figure 6.—PGL 2078: $HfH2+C_{36}H_{74}$+Mo gamma spectra after 4-h exposure; 60-min counting interval. Green: PGL 2078 results; red: cave background.

### 3.1.2 Control Tests

Tests performed with hydrogenated materials (instead of deuterated) $ErH_3+C_{36}H_{74}$+Mo and $HfH_2+C_{36}H_{74}$+Mo showed that no new spectral lines were created from exposure. Figure 6 presents example gamma results for both the exposed materials (PGL 2078) and the cave background, showing that only background lines are present. It is noted that the bismuth lines ($^{214}$Bi) normally associated with the background and the uranium decay series is slightly elevated in the specimen postexposure scans above background. This may possibly be due to the slightly elevated bismuth content found in the molybdenum prescanned materials.

Tests were also performed with non-gas-loaded erbium and hafnium, which also showed no new spectral lines were created, as expected.

### 3.1.3 Half-Life

Tables VI and VII present half-life results for two sets of samples: $ErD_{2.8}C_{36}D_{74}$+Mo (PGL 2150 to 2153) and $HfD_2+C_{36}D_{74}$+Mo (PGL 2142 to 2145), respectively. Each table shows the net area counts (determined by GammaVision® 7) for each gamma peak and corresponding suspected radioisotopes. Repeat 15-min gamma scans were performed for the shorter half-life radioisotopes. Repeat 60-min scans were done spaced over the course of days and weeks for the multiday half-life radioisotopes. Scans were compared to other comparable-length scans from which computed half-lives of each of the gamma lines noted were calculated using standard techniques. Percent differences from the published values were calculated using the average experimental half-lives, as noted. Generally the half-lives ranged from excellent to fair agreement. The half-life of $^{99m}$Tc was not determined for the PGL 2150 to 2153 data sets because of an issue with scan intervals used. In addition, since there is an overlap of the 140.51 keV line is “double-counted,” and accurate half-lives cannot be calculated for these two radioisotopes. As Table V shows, this is also reflected as an error in the $^{99}$Mo calculated half-life, which is compromised by the $^{99m}$Tc decay.

The half-life determination s are also complicated by the radioisotope half-life and the irradiation time. If the irradiation time is long compared to the half-life, then the parent and daughter will be in secular equilibrium, and the irradiation time end can be chosen as the starting point for the half-life determination. Similarly, if the half-life is long compared to the irradiation time, the irradiation time end again can be chosen. However, if the half-life is comparable to the irradiation time, then the half-life is complicated by continual, and simultaneous, radioisotope production and decay. Furthermore, this assumes a linear production rate of radioisotopes.

TABLE VI.—HALF-LIFE DETERMINATION FOR RADIOISOTOPES FOUND WHEN EXPOSING $ErD_{2.8}+C_{36}D_{74}+Mo$ (PGL 2150 TO 2153) DETERMINED FROM GAMMA LINES MEASURED AFTER EXPOSURE[a]
[Accepted gamma energy data from Lawrence Berkeley National Laboratory (Ref. 11).]

| Radioisotope/ published half-life | | | Observed gamma lines[b] | | | | Experiment average half-life | Experiment vs. published, % difference |
|---|---|---|---|---|---|---|---|---|
| | | | Highest intense line | Second intense line | Third intense line | Fourth intense line | | |
| $^{101}$Tc/ 14.22 min | | | 306.83 keV | 545.05 keV | | | | |
| | Net area counts | Scan 1 | 1768 | 77 | | | | |
| | | Scan 2 | 880 | 43 | | | | |
| | Calculated half-life, min | | 19 | 23 | | | 21 | |
| | Net area counts | Scan 2 | 880 | 43 | | | | |
| | | Scan 3 | 469 | 8 | | | | |
| | Calculated half-life, min | | 18.8 | 7 | | | 13 | |
| | Net area counts | Scan 1 | 1768 | 77 | | | | |
| | | Scan 3 | 469 | 8 | | | | |
| | Calculated half-life, min | | 19 | 11.1 | | | 15 | |
| | | | | | | | **16 min** | **15%** |
| $^{101}$Mo/ 14.61 min | | | 590.93 keV | 191.94 keV | 1012.53 keV | 505.97 keV | | |
| | Net area counts | Scan 1 | 79 | 141 | 49 | 46 | | |
| | | Scan 2 | 23 | 47 | 7 | 16 | | |
| | Calculated half-life, min | | 11 | 12 | 7 | 13 | 11 | |
| | Net area counts | Scan 1 | 79 | 141 | 49 | 46 | | |
| | | Scan 3 | 15 | 35 | 12 | 0 | | |
| | Calculated half-life, min | | 21.3 | 30.9 | 18.7 | N/C | 24 | |
| | Net area counts | Scan 2 | 23 | 47 | 7 | 16 | | |
| | | Scan 3 | 15 | 35 | 12 | 0 | | |
| | Calculated half-life, min | | 27.7 | 40.1 | N/C | N/C | 34 | |
| | | | | | | | **23 min** | **57%** |
| $^{163}$Er/ 75 min | | | 47.55 keV | 46.7 keV | 53.82 keV | 55.56 keV | | |
| | Net area counts | Scan 1 | 368 | 248 | 126 | 146 | | |
| | | Scan 3 | 286 | 133 | 103 | 113 | | |
| | Calculated half-life, min | | 100 | 40 | 125 | 98 | 91 | |
| | Net area counts | Scan 1 | 368 | 248 | 126 | 146 | | |
| | | Scan 2 | 285 | 196 | 98 | 137 | | |
| | Calculated half-life, min | | 52 | 57 | 53 | 210 | 93 | |
| | Net area counts | Scan 2 | 285 | 196 | 98 | 137 | | |
| | | Scan 3 | 286 | 133 | 103 | 113 | | |
| | Calculated half-life, min | | N/C | 31 | N/C | 61 | 46 | |
| | | | | | | | **77 min** | **2%** |
| $^{171}$Er/ 7.52 h | | | 308.33 keV | 295.94 keV | 50.74 keV | 111.66 keV | | |
| | Net area counts | Scan 2 | 2963 | 1472 | 232 | 515 | | |
| | | Scan 3 | 2896 | 1363 | 215 | 529 | | |
| | Calculated half-life, h | | 8.62 | 2.56 | 2.59 | N/C | 5 | |
| | Net area counts | Scan 1 | 3099 | 1447 | 218 | 522 | | |
| | | Scan 3 | 2896 | 1363 | 215 | 529 | | |
| | Calculated half-life, h | | 6.19 | 7.01 | 30.26 | N/C | 14 | |
| | Net area counts | Scan 1 | 3099 | 1447 | 218 | 522 | | |
| | | Scan 2 | 2963 | 1472 | 232 | 515 | | |
| | Calculated half-life, h | | 4.95 | N/C | N/C | 16.46 | 11 | |
| | | | | | | | **10 h** | **33%** |
| $^{99}$Mo/ 65.94 h | | | 140.5 keV | 739.5 keV | 181.063 keV | | | |
| | Net area counts | Scan 5 | 1423 | 205 | 176 | | | |
| | | Scan 8 | 1112 | 67 | 102 | | | |
| | Calculated half-life, h | | 271.19 | 59.8 | 122.59 | | 91.2 | |
| | | | | | | | **91 h** | **38%** |
| $^{99m}$Tc/ 6.01 h | | | 140.5 keV | | | | | NA |
| | Note: Scan intervals performed did not allow half-life assessment of the $^{99m}$Tc isotope. | | | | | | | |

[a]Test date: 30-Oct.-2015; end time = 12:15; specimen mass = 56.0010 g; scanned mass = 56.0040 g for scans 1 to 4; scanned mass = 54.6615 g for scans 5 to 10. Specimens weighed with the Mettler Toledo scale; Model: MS105DU and accurate to five decimals places. Scale was set to display mass up to four decimal places.

[b]N/C indicates values were not calculated.

TABLE VII.—HALF-LIFE DETERMINATION FOR RADIOISOTOPES FOUND WHEN EXPOSING $HfD_2$+$C_{36}D_{74}$+Mo (PGL 2142 TO 2145) DETERMINED FROM GAMMA LINES MEASURED AFTER EXPOSURE
[Accepted gamma energy data from Lawrence Berkeley National Laboratory (Ref. 11).].

| Radioisotope published half-life | | | Observed gamma lines | | | | Experiment average half-life | Experiment vs. published, % difference |
|---|---|---|---|---|---|---|---|---|
| | | | Highest intense line | Second intense line | Third intense line | Fourth intense line | | |
| $^{101}$Tc/ 14.22 min | | | 306.83 keV | 545.05 keV | | | | |
| | Net area counts | Scan 1 | 4848 | 183 | | | | |
| | | Scan 2 | 1194 | 85 | | | | |
| | Calculated half-life, min | | 19 | 35 | | | 27 | |
| | Net area counts | Scan 2 | 1194 | 85 | | | | |
| | | Scan 3 | 364 | 0 | | | | |
| | Calculated half-life, min | | 17 | N/C | | | 17 | |
| | Net area counts | Scan 1 | 4848 | 183 | | | | |
| | | Scan 3 | 364 | 0 | | | | |
| | Calculated half-life, min | | 18.2 | N/C | | | 18 | |
| | | | | | | | **21 min** | **45%** |
| $^{101}$Mo/ 14.61 min | | | 590.93 keV | 191.94 keV | 1012.53 keV | 505.97 keV | | |
| | Net area counts | Scan 1 | 228 | 332 | 121 | 168 | | |
| | | Scan 2 | 41 | 71 | 6 | 20 | | |
| | Calculated half-life, min | | 16 | 18 | 9 | 13 | 14 | |
| | Net area counts | Scan 1 | 228 | 332 | 121 | 168 | | |
| | | Scan 3 | 17 | 22 | 6 | 44 | | |
| | Calculated half-life, min | | 18.2 | 17.4 | 15.7 | 35 | 22 | |
| | Net area counts | Scan 2 | 41 | 71 | 6 | 20 | | |
| | | Scan 3 | 17 | 22 | 6 | 44 | | |
| | Calculated half-life, min | | 23 | 17 | N/C | N/C | 20 | |
| | | | | | | | **19 min** | **28%** |
| $^{180m}$Hf/ 5.5 h | | | 332.27 keV | 443.14 keV | 215.25 keV | | | |
| | Net area counts | Scan 1 | 427 | 338 | 339 | | | |
| | | Scan 3 | 331 | 294 | 263 | | | |
| | Calculated half-life, h | | 3.09 | 5.65 | 3.1 | | 4 | |
| | Net area counts | Scan 2 | 402 | 332 | 276 | | | |
| | | Scan 3 | 331 | 294 | 263 | | | |
| | Calculated half-life, h | | 1.73 | 2.76 | 6.96 | | 4 | |
| | Net area counts | Scan 1 | 427 | 338 | 339 | | | |
| | | Scan 2 | 402 | 332 | 276 | | | |
| | Calculated half-life, h | | 7.49 | 25.22 | 2.2 | | 12 | |
| | | | | | | | **6 h** | **21%** |
| $^{181}$Hf/ 42.39 days | | | 482.00 keV | 132.94 keV | 345.83 keV | | | |
| | Net area counts | Scan 4 | 506 | 252 | 116 | | | |
| | | Scan 11 | 335 | 156 | 80 | | | |
| | Calculated half-life, days | | 53.31 | 45.84 | 59.17 | | 53 | |
| | Net area counts | Scan #4 | 506 | 252 | 116 | | | |
| | | Scan #8 | 392 | 125 | 72 | | | |
| | Calculated half-life, days | | 64.43 | 23.46 | 34.49 | | 41 | |
| | Net area counts | Scan #4 | 506 | 252 | 116 | | | |
| | | Scan #9 | 324 | 97 | 80 | | | |
| | Calculated half-life, days | | 46.44 | 21.69 | 55.72 | | 41 | |
| | | | | | | | **45 days** | **6%** |
| $^{99}$Mo/ 65.94 h | | | 140.50 keV | 739.50 keV | | | | |
| | Net area counts | Scan 4 | 2869 | 221 | | | | |
| | | Scan 6 | 1331 | 100 | | | | |
| | Calculated half-life, h | | 63.66 | 61.66 | | | **63** | |
| | | | | | | | **63 h** | **–5%** |
| $^{99m}$Tc/ 6.01 h | | | 140.50 keV | | | | | |
| | Net area counts | Scan 2 | 716 | | | | | |
| | | Scan 3 | 654 | | | | | |
| | Calculated half-life, h | | 6 | | | | 6 | |
| | | | | | | | **6 h** | **–0.2%** |

[a]Test date: 28-Oct.-2015; end time: 12:23, specimen mass = 56.0082 g; Scanned mass = 56.0082 g for scans 1 to 3; scanned mass = 54.72816 g for scans 4 to 14. Specimens weighed with the Mettler Toledo scale; Model: MS105DU and accurate to five decimals places. Scale was set to display mass up to four decimal places.
[b]N/C indicates values were not calculated.

## 3.2 Beta Measurements

Both gas proportional alpha/beta and liquid scintillation instruments were used to quantify activity of material before and after gamma exposure.

### 3.2.1 Gas Proportional Alpha/Beta Counting

All as-received materials were alpha/beta counted before exposure, and none were found to have any activity above background. Within approximately 45 min after beam exposure, a portion (nom. 0.5 g) of the sample was split off for alpha/beta counting using the Tennelec gas proportional counter (protocol: 10 min alpha, then 10 min beta). None of the specimens exhibited alpha activity above background. However, positive beta activity was measured for all of the deuterated samples. Table VIII shows the beta activity of specimens measured in this study. The deuterated samples all exhibited net counts of beta activity multiple times background (ranging from 5× to 190× background). The hydrogenated samples showed no activity above background. The bare non-gas-loaded Er+Mo material showed no activity; the Hf+Mo material showed net counts barely above background. It is noted that the postexposure deuterated erbium samples had greater beta activity than the deuterated hafnium materials, both on an absolute and per gram basis.

### 3.2.2 Beta Scintillation Counting

Beta scintillation counting was performed to ascertain the beta energy spectrum. Unfortunately the specimens loaded with D-paraffin showed relatively high LUMEX numbers. The high LUMEX prevented obtaining conclusive results, so their results are not reported.

## 3.3 Neutron Activity

Evidence of neutron activity was tracked via three methods: bubble detectors, neutron activation materials, and solid-state nuclear track detectors.

TABLE VIII.—GAS PROPORTIONAL BETA COUNT RESULTS AFTER EXPOSURE FOR EACH OF THE SPECIMENS TESTED

| Specimen | Counted specimen mass, g | Beta activity | | |
|---|---|---|---|---|
| | | Background counts, cpm | Net counts, cpm | Net counts per gram, cpm/g |
| $ErD_x+C_{36}D_{74}+Mo$ | | | | |
| PGL 2028a | 9.33418 | 2 | 53 | 6 |
| PGL 2029aa | 9.00862 | 2 | 71 | 8 |
| PGL 2050 | 14.0000 | 1 | 97 | 7 |
| PGL 2051 | 14.0000 | 1 | 94 | 7 |
| PGL 2052 | 8.95773 | 1 | 191 | 21 |
| PGL 2053 | 8.18824 | 1 | 124 | 15 |
| $HfD_2+C_{36}D_{74}+Mo$ | | | | |
| PGL 2034 | 9.32799 | 3 | 17 | 2 |
| PGL 2035 | 9.65982 | 3 | 19 | 2 |
| PGL 2068 | 9.06773 | 3 | 15 | 2 |
| PGL 2069 | 9.20213 | 3 | 19 | 2 |
| PGL 2070 | 10.39806 | 3 | 15 | 1 |
| PGL 2071 | 10.6346 | 3 | 15 | 1 |
| $ErH_3+C_{36}H_{74}+Mo$ | | | | |
| PGL 2064 | 9.08073 | 2 | 1 | 0 |
| PGL 2065 | 9.9614 | 2 | 0 | 0 |
| $HfH_2+C_{36}H_{74}+Mo$ | | | | |
| PGL 2077 | 8.88926 | 2 | 1 | 0 |
| PGL 2078 | 8.73642 | 2 | 1 | 0 |
| Er+Mo | | | | |
| PGL 2061 | 4.15695 | 1 | 1 | 0 |
| Hf+Mo | | | | |
| PGL 2060 | 11.4268 | 1 | 4 | 0 |

### 3.3.1 Bubble Detectors

Bubble Detectors, manufactured by Bubble Detector Technology Industries in Chalk River, Ontario, Canada, are reusable personal neutron dosimeter that are insensitive to charged particles and gamma rays. These detectors have been used for decades terrestrially in aerospace and in space. When a neutron interacts with the detector material it produces a superheated nucleation site resulting in a visually countable bubble. However, bubble detectors have a neutron detection efficiency of $10^{-4}$.

Table IX provides the bubble counts for the detector placed on either the top of or lateral to the LINAC head, out of the direct exposure of the LINAC beam. There were a significant number of bubbles in the cases where deuterated specimens were exposed to the beam. For tests without deuterium loading, there were no bubbles recorded above background levels (i.e., only 1 to 2 over a 4-h interval). Note once the bubble count reached nominally 100 each, counting required a microscope with a special stage to manipulate the detector, which was only available for the later tests starting on August 6, 2015.

### 3.3.2 Neutron Activation Materials

When deuterium was included in the test run, the Gd and Cd neutron activation witness materials activated in all cases. Postexposure gamma lines were observed for $^{159}$Gd , as a result of the reaction: $^{158}$Gd (n,γ) $^{159}$Gd. Postexposure gamma lines were observed for $^{115}$Cd, as a result of the reaction: $^{114}$Cd (n,γ) $^{115}$Cd. Since Gd and Cd have huge cross sections for thermal neutrons, this would indicate that thermal neutrons must have been created during exposure. Anderson reports in Ref. 20 that $^{111m}$Cd is produced with 1190-keV gamma stimulation, so this line was not attributed to neutron activity when observed in the posttest analysis.

TABLE IX.—BUBBLE DETECTOR COUNTS FOR PGL SAMPLES NOTED

| Test specimen ID | Test date | Exposure duration, hr | Bubble Detector serial number, location | Bubble Detector sensitivity, bub/mRem | Counts |
|---|---|---|---|---|---|
| $ErD_x+C_{36}D_{74}+Mo$ | | | | | |
| PGL 2028a | 28-Jul-15 | 4 | 14162329, top of LINAC | 12.0 | >100 |
| PGL 2029aa | | 4 | 14120319, top of LINAC | 9.7 | ~50 |
| PGL 2050 | 3-Aug-15 | 4 | 14162329, top of LINAC | 12.0 | >100 |
| PGL 2051 | | 4 | 14122516, top of LINAC | 9.9 | >100 |
| PGL 2052 | | | | | |
| PGL 2053 | | | | | |
| PGL 2150 | 30-Oct-15 | 6 | 15114323, wave guide | 23 | 55 |
| PGL 2151 | | 6 | 15114326, top of LINAC | 22 | 61 |
| PGL 2152 | | | | | |
| PGL 2153 | | | | | |
| $HfD_2+C_{36}D_{74}+Mo$ | | | | | |
| PGL 2034 | 29-Jul-15 | 4 | 14162537, top of LINAC | 12.0 | >100 |
| PGL 2035 | | 4 | 14162516, top of LINAC | 9.9 | >100 |
| PGL 2068 | 6-Aug-15 | 4 | 14162527, top of LINAC | 11.0 | 448 (via microscope) |
| PGL 2069 | | 4 | 14162537, top of LINAC | 12.0 | 45 |
| PGL 2070 | | 4 | 14162329, side | 12 | 67 |
| PGL 2071 | | 4 | 13144401, cooling hole | ? | |
| PGL 2142 | 28-Oct-15 | 6 | 15113338, top of LINAC | 22 | 131 |
| PGL 2143 | | 6 | 15113420, top of LINAC | 23 | 186 |
| PGL 2144 | | 6 | | | |
| PGL 2145 | | 6 | | | |
| PGL 2228 | 15-Dec-15 | 6 | 15114119, wave guide | 21 | 204 |
| PGL 2229 | | 6 | 15114127, top of LINAC | 22 | 183 |
| PGL 2230 | | 6 | 15113335, inside top cave | 22 | 89 |
| PGL 2231 | | 6 | 15113353, side | 23 | 19 |
| $Er\text{-}H_3$ + H-para + Mo | | | | | |
| PGL 2064 | 5-Aug-15 | 4 | 14162555, top of LINAC | 12.0 | 1 |
| PGL 2065 | | 4 | 14161225, top of LINAC | 12.0 | 0 |
| $ErH_3+C_{36}H_{74}+Mo$ | | | | | |
| PGL 2077 | 7-Aug-15 | 4 | 14163552, top of LINAC | 10.0 | 0 |
| PGL 2078 | | 4 | 14162516, top of LINAC | 9.9 | 0 |
| Er+Mo | | | | | |
| PGL 2061 | 4-Aug-15 | 4 | 14163552, top of LINAC | 10.0 | 1 |
| Hf+Mo | | | | | |
| PGL 2060 | 4-Aug-15 | 4 | 14162555, top of LINAC | 12.0 | 0 |

Indium was also used as a neutron witness material. Two radioisotopes of indium were observed: $^{115m}$In, from gamma rays and fast neutrons, and $^{116m}$In, from slow neutrons. $^{115m}$In will activate via either a gamma-induced reaction, $^{115}$In (γ, γ')$^{115m}$In, or an inelastic neutron interaction, $^{115}$In(n, n')$^{115m}$In. Collins et al. (Ref. 21) reports that $^{115m}$In is produced with 1078-keV gamma stimulation. $^{116m}$In activation is caused by inelastic neutron capture: $^{115}$In(n, γ)$^{116m}$In, having a capture cross section resonance at 1.457 eV (epithermal). The $^{116}$In 14-s half-life is unobservable after the LINAC irradiation, whereas $^{116m}$In has a 54-min half-life, making it accessible after irradiation. Because the LINAC produces gamma rays up to 1.95 MeV, the gamma stimulation level of $^{115m}$In was achieved, and the gamma scans indicating the presence of $^{115m}$In are not reliable enough to be conclusive for fast neutrons at this point. Designing a better tungsten/lead cave, on top of the LINAC but close to the sample, to shield out all gamma rays is being accomplished, and subsequent experimental work should include more reliable data involving indium as a neutron witness material.

### 3.3.3 Solid-State Nuclear Track Detectors

Solid-state nuclear track detectors (SSNTDs, the CR–39 detectors) have been used for decades. CR–39 was developed in the 1970s (Ref. 22) and was subsequently used for space (Ref. 23) and inertial fusion particle spectroscopy (Ref. 24). CR–39 can also be used to detect neutrons by various neutron inelastic and elastic interactions with the hydrogen, carbon, and oxygen constituents of CR–39 resulting in recoils, or fast-neutron energy-dependent spallation and fragmentation of the carbon and oxygen atoms in the CR–39. These leave tracks that can be etched to microscopically observable dimensions. However, the fast-neutron detection efficiency ranges from $10^{-4}$ to $10^{-5}$ tracks/neutron (Ref. 25). CR–39 has a fast-neutron detection threshold of about 100 to 200 keV, below which energy no etched track is observable.

CR–39 detectors were analyzed for several tests where deuterated materials were exposed (PGL 2034 and 2035, $HfD_2$+$C_{36}D_{74}$+Mo) and for the corresponding hydrogenated baseline specimen (PGL 2060, Hf without D loading + Mo). There were no tracks above background for the exposure case of PGL 2060, which had no deuterium loading. For the exposure case with PGL 2034 and 2035 (with D loading), CR–39 detectors were used to measure and compare the track count on the CR–39 near the exposed samples (CR–39: SN 569532 and 569533) and to compare them to background detectors (CR–39: SN 5693470 and 5693471) placed in a control room a distance from the LINAC site. For reference, the CR–39 detectors were placed with the numbered side facing away from the LINAC and the obverse side facing towards it. Tracks were counted and lengths measured for the LINAC-mounted detector front (numbered) and back (obverse) sides, and the results are plotted in Figure 7. Two CR–39 detectors on the LINAC showed counts in excess of 2 to 3 times the background, with SN 569532 having 175 tracks against a background of 54 tracks. The 1-mm-thick CR–39 acts as a neutron radiator, increasing two-fold the neutron detection efficiency.

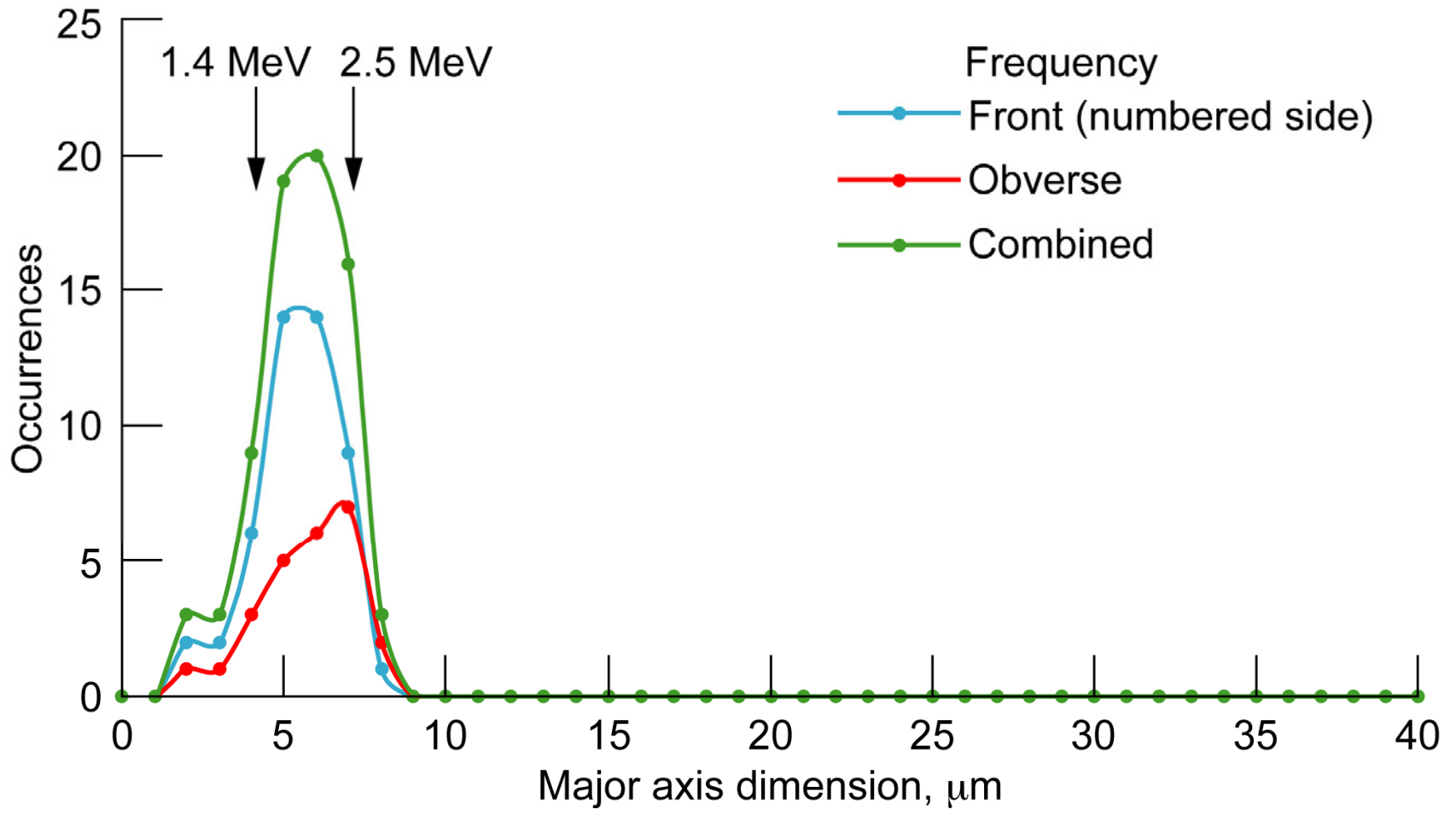

Figure 7.—Track major axes of CR–39 (SN 569532), including front (numbered) and back (obverse) side, and total counts plotted with reference to approximate neutron energy calibration data from International Atomic Energy Agency (IAEA) normalized data (Ref. 26). Data binned in 1-μm increments.

Neutron energy calibration was done by irradiating the same type of Fukuvi CR–39 with an unmoderated $^{252}$Cf neutron source, etching it, and scanning it with a Track Analysis Systems Ltd. (TASL) scanner that provides major and minor axis measurements. The track major axis count frequency was normalized to a standard IAEA $^{252}$Cf neutron spontaneous fission source (Ref. 26) (1 to >9 MeV). The energies noted in the plot are estimated, ±500 keV, from the $^{252}$Cf IAEA normalization.

Based on the above comparison, it appears that the track lengths in the LINAC-based CR–39 are consistent with neutron energies in the range between 1.4 and 2.5 MeV neutron energies. Note the CR–39 "front" side energy plot peaks just under 2±0.5 MeV.

At the operating gamma end-point energy of 1.95 MeV there should be no photodissociated deuterium and hence no photoneutrons. Even at the highest endpoint at which this LINAC is rated, 2.4 MeV, these high-energy neutrons would be precluded. For instance, if the machine were operating out of standard control (which it was not), the peak energy the machine could produce is 2.4 MeV. Using the value of 2.226 MeV for deuterium photodissociation coupled with the observation that half the kinetic energy after photodissociation would go to the proton and half to the neutron, the maximum achievable neutron energy would be 0.087 MeV. This is clearly below the ranges of neutron energies found and below the CR–39 fast neutron detection threshold of 144 keV (Ref. 25). Additionally, the minimum neutron energy the Bubble Detector would indicate is in the 100 to 200 keV range.

On a later set of tests performed with a 1.95-MeV exposure of samples PGL 2189 to PGL 2192 consisting of $HfD_2$, $C_{36}D_{74}$, and Mo, a set of CR–39 detectors was placed on the LINAC head (CR–39: SN 5694012 and 5694013). To avoid photon damage, the CR–39 detector was placed approximately 18 cm away from the braking target. Furthermore there was about 5 cm of lead between the beam braking target, specimens, and the CR–39 detector. The background CR–39 detector was placed in an external control room approximately 30 m from the LINAC and separated from the reaction by ~0.1 m of concrete and about 1 to 2 m of dirt overburden covering the bunker. Consequently, only background neutrons should reach the CR–39 detector.

Upon examination of the LINAC CR–39 detector (SN 5694013), a triple track was found. A triple track is indicative of a neutron with an energy >10 MeV causing a $^{12}C_6(n,n')3\alpha$ reaction, resulting in the disintegration of a carbon atom in the CR–39 (Ref. 27). This is additional evidence that high-energy neutrons were present. Although the neutron will not leave a track, up to three alpha particles arising from a common point will leave ionization trails. It is noted that the relative efficiency of causing a triple track is in the range of $10^{-4}$ to $10^{-5}$. The two photomicrographs provided in Figure 8 show the triple track taken at two different depths of focus. The image in Figure 8(a) is on the CR–39 surface, and the image in Figure 8(b) is focused deeper within. The length of each of the alpha tracks along with the $^{12}C_6$ binding energy gives the energy of the incident neutron.

Figure 9 shows an example of a photomicrograph of a triple track produced by a deuterium-tritium (D-T) fusion generator. In several years of CR–39 use, and in the several hundred detectors examined, the authors have never seen a triple track in the background, reinforcing that the triple track was induced by the experiment. It is noted that the beam energies used herein are insufficient to cause neutrons from deuterium photodissociation or photoneutron production from Hf or Mo.

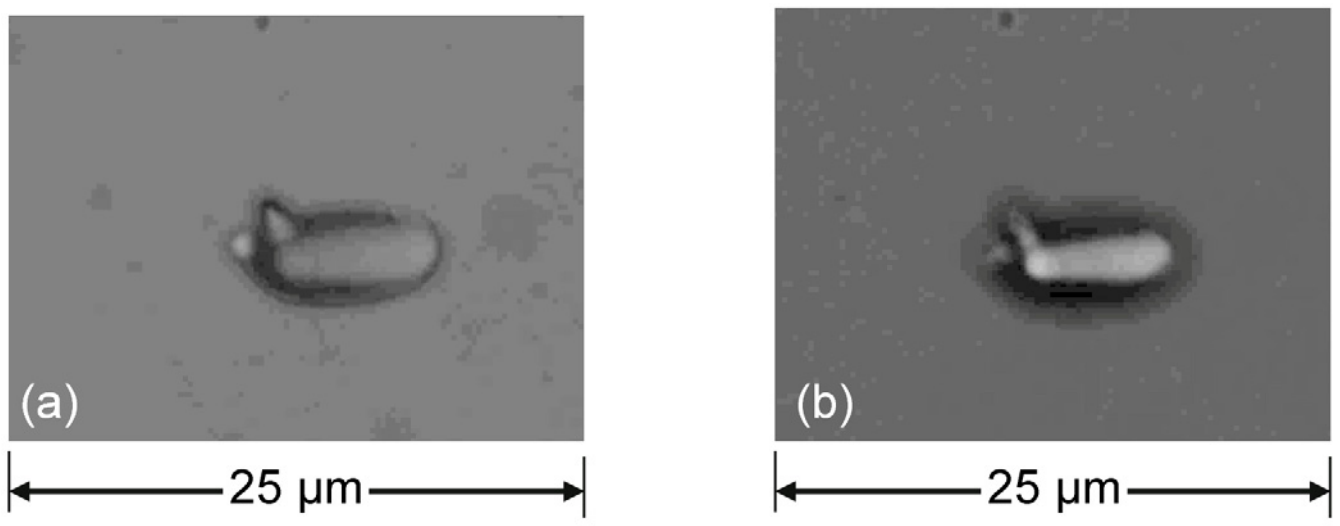

Figure 8.—Photomicrographs of triple track found in LINAC solid-state nuclear track detector CR–39 at two focal depths. (a) At surface of etched CR–39 detector. (b) Deeper within triple track.

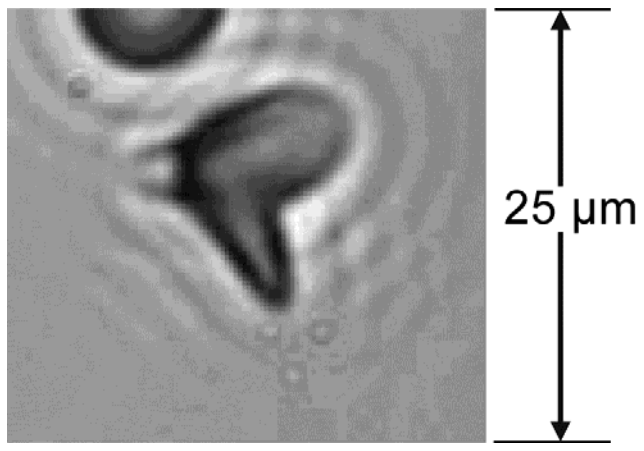

Figure 9.—Photomicrograph of typical triple track found in solid-state nuclear track detector CR–39 when exposed to deuterium-tritium (D-T) fusion generator.

# 4.0 Discussion

Considering the neutron activity data collected and the conditions of the experiment, several mechanisms are discussed.

## 4.1 Neutron Energies

Based on the tests performed and subsequent analysis of the data, it is clear that a range of neutron energies were created, only when irradiating samples containing deuterium. The $^{115}$Cd and $^{159}$Gd neutron activation materials indicate that thermal neutrons were present. When irradiating samples that were hydrogenated or were unloaded, no neutrons were measured by the cadmium material. The Bubble Detectors indicated neutrons with energies $\geq$100 keV were present. The CR–39 detectors indicated that appreciable neutrons in the 1.4 to 2.5 MeV range were present. The detector's minimum threshold for detection is 144 keV. On several detectors triple tracks were found, indicating neutron energies 10 MeV or greater.

## 4.2 Mechanisms Considered

There are several plausible mechanisms for the observed neutron activations:

(1) Deuteron photodissociation: Photodissociation was carefully considered, as this is an obvious source of neutrons. However, the beam end-point energy was shown to be below 2226 keV (deuterium photodissociation threshold) by the water tank ionization chamber method. Based on this method, the beam was operated below the photodissociation threshold, and therefore is excluded from being the source of neutrons.

(2) Electron capture: Parent metal capture of electrons was also considered. Certain radioisotopes can be formed via electron capture. However, if electron capture would have occurred, then these spectral lines would have been obvious when exposing either the hydrogen-loaded materials or the bare, non-gas-loaded metals. No such activation was observed. Activation only occurred with those reactants containing deuterium.

(3) Gamma-n processes: Photoneutron reactions were also considered. Classic photoneutrons are not the cause of the activation of the Hf, Er, or Mo in the deuterated cases. First, the gamma energy is several MeV too low to cause photoneutrons in these materials. Second, if photoneutrons were at work they would have occurred in both the hydrogenated and non-gas-loaded materials, but these reactants did not show activation.

(4) Gamma-metastable processes: As mentioned, $^{180m}$Hf was created in the deuterated cases where hafnium was included. It was postulated that $^{180m}$Hf may have been created by the gamma energy exciting $^{180}$Hf to the metastable condition. However this is not the mechanism, as again it would have occurred for either the H-loaded or non-gas-loaded hafnium specimens exposed to the beam under the exact same conditions, but these metastable radioisotopes did not appear in those instances.

(5) Mossbauer effect: Upon gamma ray excitation of nuclei that are embedded in the lattice, and if the recoil energy is not sufficient to generate a phonon, there is the finite probability that the entire crystal recoils, rather than an individual nucleus, resulting in essentially a recoilless emission of gamma-photon. This process is called the Mossbauer effect. However, it is well known that the Mossbauer effect results in milli-electron-volt energy levels, which are irrelevant in comparison to the energies being considered in this study.

(6) Three-body interactions: Theoretically, one can postulate an event in which fluorescent photons arrive at a target nucleus also being exposed with a photon having energy very close to photodissociation, together exceeding the photodissociation energy. Aside from the typical exceedingly low probability of a three-body interaction, the highest fluorescent photons from the present experiments are in the 50 to 60 keV range.

(7) Contribution from external sources: Activation was only found when D-fuel was included in the specimens. This would argue strongly against specimen contamination from external sources. Regarding potential contamination, the gamma energy spectral lines were specific for the radioisotopes identified in the paper and were not those for lead, thorium, or radon radioisotopes, and were further confirmed with half-life determination.

The Standard Model of particle physics, incorporating the electromagnetic, weak, and strong forces may provide for other possible mechanisms that account for the reactions observed herein. However, discussion of those is beyond the scope of the current experimental work.

## 4.3 Future Work

Because of the limitations in the ability of measuring beam end-point energy with the available equipment, the authors strongly recommend that these tests be repeated using equipment where the beam end-point energy is specified to a high degree of confidence (i.e., 5 sigma or greater) and is below the deuterium photodissociation threshold. Testing in a facility with the appropriate means of end-point energy verification is currently being sought.

# 5.0 Summary of Results

Exposure of highly deuterated materials to a low-energy (nom. 2-MeV) photon beam resulted in nuclear activity of both the parent metals of hafnium and erbium and a witness material (molybdenum) mixed with the reactants:

1. Gamma spectrum: Gamma spectral analysis of all of the deuterated materials $ErD_x$+$C_{36}D_{74}$+Mo and the $HfD_2$+$C_{36}D_{74}$+Mo materials showed that nuclear processes had occurred during exposure as shown by unique gamma signatures. For the deuterated erbium specimens, posttest gamma spectra showed evidence of radioisotopes of erbium ($^{163}$Er and $^{171}$Er) and molybdenum ($^{99}$Mo and $^{101}$Mo) and by beta decay, technetium ($^{99m}$Tc and $^{101}$Tc). For the deuterated hafnium specimens, posttest gamma spectra showed evidence of radioisotopes of hafnium ($^{180m}$Hf and $^{181}$Hf) and of molybdenum ($^{99}$Mo and $^{101}$Mo) and by beta decay, technetium ($^{99m}$Tc and $^{101}$Tc). In contrast, when either the hydrogenated or non-gas-loaded erbium or hafnium materials were exposed to the gamma flux, the gamma spectra revealed no new radioisotopes. The gamma spectra peaks showed only background decay lines. Further study is required to determine the mechanism by which the neutron activity is happening.
2. Alpha/beta results: The alpha/beta counting performed showed that there was no activity above background prior to exposure. However, the deuterated samples all exhibited net counts of beta activity multiple times background (5× to 190× background) after exposure. The hydrogenated samples showed no activity above background after exposure.
3. Neutron energy: When deuterated materials were exposed, neutron activity was observed. The cadmium and gadolinium witness materials (placed in vials adjacent to the primary vials) showed evidence of thermal energy neutrons. The Bubble Detector dosimeters showed clear evidence that when deuterated specimens were exposed to the beam, a significant flux of neutrons were created in the >100- to 200-keV energy range. The CR–39 polycarbonate detectors placed above the LINAC head showed clear evidence of fast neutrons (1.4 to 2.5 MeV) during fueled shots. Also, in some fueled experiments the CR–39 detectors recorded triple tracks, indicating neutrons with energies ≥10 MeV.

## Acknowledgments

The authors gratefully acknowledge the assistance of many people that supported this effort. LINAC Test operators including Brian Jones, Paul Stout, Chris Maloney, Nick Connelly, John Zang, and Mark Worley; Operations Management: Jerry Hill, Dave Stringer, Tom Keating, Wes Sallee, and Jeremiah Folds; Nuclear Diagnostics and Health Physics support: Karen Novak, Cathy Jensen, Becky Johannsen, Megan Sherman and Dave Lyera; Materials analyses: Dr. David Ellis, Dr. Kathy Chuang, Dr. Ivan Locci, and Dan Scheiman; Materials loading/ and sample preparation: Mr. Frank Lynch, Dr. Dave Ellis, and Mr. Phillip Smith; Dr. Richard Kroeger for his expertise in interpreting gamma end-point energy results; Tracy Kamm, and Dave Hervol for database and record support; LINAC equipment support: Justin and Mike Summers, Acceletronics; Dr. Pamela Mosier- Boss, Amy Rankin, Victoria Leist, and Emily Martin for CR-39 measurement and analysis. We are also grateful for Gus Fralick’s comments on the manuscript and Laura Becker’s patient editorial attention. This work was supported by NASA's Planetary Science Division, Science Mission Directorate.